\documentstyle[11pt,bezier]{article}
\begin{document}

\hyphenation{Schwarz-schild}

\def\scri{
\unitlength=1.00mm
\thinlines
\begin{picture}(3.5,2.5)(3,3.8)
\put(4.9,5.12){\makebox(0,0)[cc]{$\cal J$}}
\bezier{20}(6.27,5.87)(3.93,4.60)(4.23,5.73)
\end{picture}}

\vbox{\baselineskip=12pt
\rightline{\large PP97--2}
\rightline{\large gr-qc/9607026}}

\bigskip\bigskip

{\Large
\centerline{Geometry of Black Holes and Multi-Black-Holes}
\centerline{in 2+1 dimensions\footnote{Lecture given at ``Raychaudhuri
session," ICGC-95 conference, Pune (India), December 1995}}}
\bigskip
\centerline{Dieter R. Brill}
\medskip
\centerline{University of Maryland, College Park, MD 20742, USA}

\bigskip
{\large \centerline{Contents}}

\hspace{1cm}I. Introduction

\hspace{1cm}II. (2+1)-Dimensional Initial Values in Stereographic Projection

\hspace{2cm}A. The single, non-rotating black hole

\hspace{2cm}B. Non-rotating multi-black-holes

\hspace{2cm}C. ``Black Hole" Universe Initial Values

\hspace{1cm}III. Time Development: Enter the Raychaudhuri Equation

\hspace{1cm}IV. Time Development in Stereographic Projection

\hspace{2cm}A. The BTZ black hole spacetime

\hspace{2cm}B. Multi-Black-Hole Spacetimes

\hspace{1cm}V. Black Holes with Angular Momentum

\hspace{1cm}VI. Analogous 3+1-Dimensional Black Holes

\hspace{1cm}VII. Conclusions

\bigskip
{\large I. Introduction}

\medskip
On this special occasion I bring greetings to Prof.~Raychaudhuri from the
Relativity Group at the University of Maryland,  a university that had the
good fortune to have Prof.~Raychaudhuri as a visitor about the time that
``his" equation was first explored and began to shed light on so many
interesting and important issues in General Relativity. I myself missed much
of this history because I did not arrive at Maryland until the 1970's, so
I am particularly grateful not to miss the celebration today, and I am indeed 
honored to participate in this happy occasion.

My topic is not directly concerned with the Raychaudhuri equation, but it is
of course not hard to establish the connection with his elegant ideas;
what would be hard is to give a talk in modern General Relativity in which
the Raychaudhuri equation would not enter in some fashion, so pervasive is
its influence in our field. 

It is perhaps surprising that gravity in 2-dimensional space, even as a
``toy model," should be of any interest, because a 
local analysis of vacuum Einstein gravity in 2+1 dimensions reveals 
little contents. Namely, Einstein's equations specify the Ricci tensor,
which in 2+1 dimensions specifies the Riemann tensor, which in turn gives
a complete local description of gravity: spacetime is flat, or has constant
curvature $\Lambda$ if there is a cosmological constant. In particular, there
is no ``Newtonian limit" of 2+1 dimensional gravity. It was therefore
surprising when, about three years ago, 2+1 dimensional black hole solutions
were discovered. Their nature is due to the {\em global} structure of 
spacetime, there is no local gravitational interaction associated with them. 

In 3+1 dimensions, exact solutions containing more than one black hole are
known only in special cases. The difficulty is that an exact multi-black-hole
solution would have to describe in one metric all the physics associated
with such configurations, including their motion in orbits about each other,
the gravitational radiation emitted, the consequent energy loss of the black
hole system, and its possible eventual collapse and merger into a single
asymptotically stationary black hole. No such complicated behavior is
expected in 2+1 dimensions, for example because there is no gravitational
radiation. Hence it is reasonable to expect exact multi-black-hole solutions
in this case.

I will show how to construct such solutions, and explore the geometry of the
spacetimes so obtained. The results are interesting not least for the
fascinating
way in which the 2+1 dimensional spacetime succeeds in imitating most of
those properties expected of 3+1 dimensional multi-black-holes that can be
achieved by the global structure. One may therefore hope that these solutions
can serve as simplified models for exploring the more difficult questions
of black hole physics, for example those connected with quantization; but
I will not attempt to show such applications here.

\bigskip
{\large II. (2+1)-Dimensional Initial Values in Stereographic Projection.}

\medskip
All 2+1 spacetimes with a given negative cosmological constant are locally 
isometric, as we have seen. One important global difference between 
such spacetimes is the topology. More complicated topologies can be 
obtained from simpler ones through identifications by certain isometries. The 
topologically simplest of these is anti-de Sitter space; it is the universal
cover of all others.

In the usual quotient space construction that implements the identification
process the isometries have no fixed points and are spacelike. This is so 
because one wants to avoid singularities or closed timelike lines in the 
quotient space. In the case of
black holes, singularities are familiar and accepted if they are hidden
behind horizons. To understand such geometrical relationships 
it is convenient to use a representation of anti-de Sitter space 
that is the extension, to spacetimes of constant negative curvature, 
of the description due to Poincar\'e
for the 2-dimensional Riemannian case. These representations are
conformal and can be considered a stereographic projection of the curved
space into the plane or into flat three-space.

We begin with the embedding of anti-de Sitter (adS) space in a flat space of 
one higher dimension. This is analogous to the standard embedding of the round
sphere in Euclidean space, but for negative curvature the embedding space
must have indefinite metric, of a signature with one more negative 
direction than that of the 
adS space itself. For 2+1 dimensional adS space, we need a four-dimensional
flat space of signature $(--++)$ and metric
$$ds^2 = -dT^2 - dU^2 + dX^2 + dY^2  \eqno(1)$$
The hyperbolic surface
$$ -T^2 - U^2 + X^2 + Y^2 = -\ell^2.\eqno(2)$$
then represents a solution
of the vacuum Einstein equations with negative cosmological constant
$\Lambda = -1/\ell^2$. 
(This embedded spacetime is periodic in the time direction;
true adS spacetime is the universal cover of this periodic spacetime,
obtained by ``unwinding" about the XY axis. Our stereographic maps will
always project less than one period, 
so that this distinction will not make any difference.)

The stereographic map of a 2-sphere embedded in the standard way in 
Euclidean 3-space can be obtained by projection from the ``North" pole
on a plane tangent at the ``South pole." For the hyperboloidal
surface of Eq (2), the
analog of the North pole can be taken to be the point $(0,\ell,0,0)$, and
the analog of
the tangent space to the South pole is the ``plane" $U=-\ell$. 
A point $X^\mu$ ($\mu = 0, \dots 3$ with $U = X^1$) in the hyperboloidal 
surface projects to a point with coordinates
$$x^\mu = {2\ell X^\mu\over U+\ell} \qquad X^\mu \neq U \eqno{\rm (3a)}$$
in three-dimensional Minkowski space. The surface metric in these 
coordinates is
$$ds^2 =\left({1\over 1-r^2}\right)^2(-dt^2 + dx^2 + dy^2) \quad {\rm where}
\quad r^2 = {-t^2 + x^2 + y^2 \over 4\ell^2},\eqno{\rm (3b)}$$
clearly a {\em conformal} map.

To picture the projection in no more than three dimensions we
consider subspaces, such as
the ``initial" spacelike surface $T=0=t$, $U>0$. This surface has zero
extrinsic curvature (it is ``totally geodesic"). Because adS space has 
constant negative curvature, the intrinsic geometry of this surface must 
likewise have constant negative curvature --- it is hyperbolic 2-space,
$H^2$. Its embedding in (1+2) Minkowski 
space is obtained by setting $T=0$, $U>0$ in Eqs (1) and (2). 
This embedding and its stereographic projection on the plane $U=-\ell\,$ 
is shown in Fig.~1. (The embedded surface is one of the two
spacelike hyperboloids of constant spacetime distance $\ell$ from the origin.)

\bigskip\medskip

The Figure shows that in the stereographic projection all points of the
hyperboloid lie within a ``limit circle" of finite radius $2\ell$. This 
representation of the two-dimensional space of constant negative curvature is 
known as the {\em Poincar\'e disk}, whose properties are very well 
understood. The stereographic projection translates any figure or theorem of 
the ``hyperbolic geometry" in the space $H^2$ of constant negative curvature 
to a figure or theorem of Euclidean geometry. Thus the figures drawn in the
Poincar\'e disk are not merely cartoons, but show geometrical relationships
as faithfully as figures of Euclidean geometry.

The metric of the Poincar\'e disk is given by Eq (3b) with $t=0$. From it we 
can deduce the shape of {\em geodesics} in the disk. Alternatively we can
characterize geodesics on the hyperboloid as intersections with planes 
through the origin. In either approach 
we find the standard result: geodesics are represented in the disk
as arcs of circles that intersect the limit circle at right
angles (as judged by the flat, Euclidean geometry of the plane). 
Two such geodesics either intersect inside the limit circle, or 
touch on the limit circle, or intersect nowhere. In the latter two cases
they are called parallel and ultraparallel, respectively.

The hyperboloid is invariant under (isochronous) Lorentz transformations,
therefore the symmetry group of $H^2$ is $SL(2,R) \sim SU(1,1)$. 
The boosts have no fixed
points on the hyperboloids; the corresponding transformations on the
Poincar\'e disk are called {\em transvections}.\footnote{Isometries of
$H^2$ can be classified into elliptic (one fixed point, we call these
rotations), parabolic (a fixed point at infinity), and hyperbolic
(the transvections).}  Because any point on the
hyperboloid can be boosted to the special point $U=\ell$, $X=Y=0=T$,
any point on the disk can be transvected to the origin. It is often
useful to make such a transvection, without loss of generality, to
get a simple representation of some geometrical feature. We can think of 
the Poincar\'e disk as a magnifying glass that gives an undistorted and
largest image only at its center, and this magnifying glass can be
centered on any point to see a true image of its immediate neighborhood.
For example,
rotations appear as Euclidean rotations of the plane if the center of
rotation is placed at the origin; and geodesics through the origin are
Euclidean straight lines. 

\bigskip
{\large A. The single, non-rotating black hole}

\medskip
The (2+1)-dimensional black holes of Ba\~nados, Teitelboim and Zanelli  
\cite{BTZ} with zero angular momentum are described by the metric
$$ds^2 = -\left(-M + (r/\ell)^2\right)dt^2 + {dr^2\over -M + (r/\ell)^2}
+ r^2 d\phi^2.\eqno(4)$$
That this is locally equivalent to the surface of Eq (2) is shown by
the embedding
$$ X = \sqrt{-\ell^2+{r^2\over M}}\, \cosh {\sqrt{M}\over\ell}t \qquad 
Y = {r\over \sqrt{M}} \sinh \sqrt{M}\phi$$
$$ T = \sqrt{-\ell^2+{r^2\over M}}\, \sinh {\sqrt{M}\over\ell}t \qquad 
U = {r\over \sqrt{M}} \cosh \sqrt{M}\phi, \eqno{(5)}$$
which transforms the metric (1) into the metric (4). Note that the
spacelike surface $t=0$ corresponds to the embedding with $T=0$ shown
in Fig.~1. The coordinate lines described by Eq (4) on this embedded
surface project on the Poincar\'e disk to the lines shown in Fig.~2.

\medskip

What distinguishes the BTZ metric globally is the identification
$\phi \equiv \phi+2\pi$. 
In the embedding of the surface $t=0=T$ a change in $\phi$ 
corresponds to a Lorentz boost that leaves the $X$-axis fixed, hence it is a
transvection in the corresponding Poincar\'e disk. The radial lines
$\phi =$ const clearly are geodesics, hence fit smoothly together. 
Thus the isometry that
leads to the initial state of a BTZ black hole is a transvection connecting
two ultraparallels, such as the two shown as thick curves in Fig.~2.
The limit circle is thereby divided into two parts, which represent the two 
asymptotically adS regions of the BTZ black hole. The ``throat" (horizon) 
of the black hole is the minimal,``vertical" geodesic between the two
thick curves.

\bigskip
{\large B. Non-rotating Multi-Black-Holes}

\medskip
Instead of identifying the two geodesics in one Poincar\'e disk as in Fig.~2,
we can use two identical copies of Fig.~2 and identify geodesics that
correspond when the two copies are superimposed (Fig.~3). 
To obtain a BTZ black hole of the same 
mass by this ``doubling," the range of $\phi$ in each disk should be half
that in the single disk construction.

\medskip\normalsize

By a similar construction \cite{B} we can obtain a space with more than two
asymptotically adS regions, describing several black holes.\footnote{Strictly
speaking, the construction gives us initial values. That these do develop
into spacetimes with horizons and black holes will be shown in Section IV.
For now we can think of the horizons as apparent horizons.}
 We start with a region between 
several mutually ultraparallel geodesics. We shall call this region the 
``original region." We make two copies and identify corresponding geodesics.
Figure 4 shows this doubling procedure for the case of three geodesics. 
Each of the three regions between
pairs of geodesics near the limit circle is isometric to the asymptotic
region of a single BTZ black hole of suitable mass $M$. To specify the mass 
parameter we recall that any pair of ultraparallels have a unique (minimal) 
geodesic segment that is normal to both. For a single BTZ black hole this
segment corresponds to the horizon, and its length is $\pi\ell\sqrt{M}$.
(This is half the horizon length because the black hole geometry is
obtained from the original region by doubling.) For the multi-black-hole
geometry the isometry of an asymptotic region to a BTZ black hole can be
extended at least to the minimal segment, so it is reasonable to associate
with each asymptotic region a mass (as measured from that region) given
by the same formula, using the minimal segment between corresponding
adjacent ultraparallels of the original region. We shall call these minimal
segments the horizon of the corresponding black hole.
In general, there will 
be as many separate, asymptotically adS regions, and hence as many 
black holes, each with its own mass parameter, as there are geodesics 
bounding the original region.

\medskip

The identifications of the Figure can also be described by a group of
isometries, so that the identified space is the quotient of $H^2$ by this
group. To find these isometries we first note that any transvection in
the Poincar\'e disk can be completely described by a geodesic segment.
(Namely, the segment specifies the plane through the origin of Fig.~1
in which its pre-image lies; the segment's endpoints specify the boost
parameter of the Lorentz transformation that leaves the plane invariant;
and the image of the Lorentz transformation is the desired transvection.)
The group of isometries is generated by the transvections specified by 
twice the normal segments of the original region that describe the horizons.
(The factor two comes about because we are doubling the original region.
For $n$ horizons only $n-1$ of the isometries are independent, Eq (6).) 

In the original region, the horizons and the parts of the ultraparallels
between them form a polygon with $2n$ sides and with all interior angles 
being right angles (Fig.~5a). The length of the horizon $h_i$ measures the
mass of the $i^{\rm th}$ black hole; the segment $s_i$ between horizons 
$h_{i-1}$ and $h_{i+1}$ is
the shortest distance between these horizons, and can be identified with
the ``distance" between the two corresponding black holes. In such a
polygon of $2n$ sides, when $2n-3$ consecutive sides are given, then the
remaining three sides are determined. In other words, we can specify
$n-1$ black holes and the distances between adjacent ones, and are
then forced to have an $n^{\rm th}$ black hole at definite distance from
the first and the $n-1^{\rm st}$ one. To express this relation algebraically
we think of the Lorentz transformation $L_i$ in the embedding picture, 
specified by the $i^{\rm th}$ side. Because the polygon closes, the
product of the $2n$ transformations must be unity,
$$\prod_i^{2n}L_i = 1\!\!1, \eqno(6)$$
a set of three conditions. For example, two horizons of mass $m_1$ and $m_2$
at a distance $d$ require a third horizon of mass $M$ such that \cite{S}
$$\cosh(\pi\sqrt{M}) = \cosh(d/\ell)\sinh(\pi\sqrt{m_1})\sinh(\pi\sqrt{m_2})
-\cosh(\pi\sqrt{m_1})\cosh(\pi\sqrt{m_2})\geq 1 \eqno(7)$$

We see that not all choices of $m_1,\, m_2$ and $d$ are possible, because the
r.h.s. must not be less than unity. For given $m_1,\,m_2$ this places a lower
limit on $d$. As $d$ approaches the lower limit, $M$ tends to zero and
its distance from the other black holes tends to infinity. The hexagon
degenerates into a pentagon with an ideal vertex at infinity (Fig.~5b). (If
$d$ is less than the lower limit, the pentagon has a finite vertex with
an angle less than a right angle. The doubling would then lead to a conical
singularity at this vertex, corresponding to a ``particle" analogous to
a cosmic string in 3+1 dimensions. Such particles are not considered
here; see however \cite{S}, \cite{D}).

\medskip

To obtain a valid polygon for our construction, the inequality of
Eq (7) must not be violated by any pair of masses. We can therefore
think of the three-particle configuration as the basic building block of
the higher ones: we cut off an asymptotic region at its horizon and 
``thread" it to another such configuration, similar to the way a plumber 
can thread together several tees to obtain a connected manifold with 
a larger number of openings (Fig.~6). Note that 
there is additional arbitrariness in the ``twist" angle between adjacent 
``tees," which makes these multi-black-holes more general than those obtained 
by doubling of a polygon.

\bigskip
{\large C. ``Black Hole" Universe Initial Values}

\medskip
The horizons of any multi-black-hole configuration are closed geodesics,
so they have vanishing extrinsic curvature. If the intrinsic geometry
(the length, hence the mass parameter) of two horizons is the same, then
they can be identified. For example, the horizons in Fig.~6b can be identified
pairwise as shown by the dotted arrows. There are then no more asymptotic
regions, we have a closed universe containing several 
``wormholes".\footnote{These are not black holes in the usual sense because
once the asymptotic region is eliminated, there is no longer a horizon around
them. They do, however, have the classic wormhole topology, and the geometry
at each wormhole there is an apparent horizon, with the geometry in its  
vicinity being the same as that near the (true) horizon of a BTZ black
hole.} The Gauss-Bonnet theorem
implies that the Euler characteristic of such a configuration must be
negative, hence the universe constructed according to Fig.~6 has the
minimum number (two) of wormholes. At each identification there is an
arbitrary angle of twist, so this type of universe is characterized by
three mass and three twist parameters, with the masses at each tee having
to satisfy the inequality (7).

Instead of eliminating all asymptotic regions we could keep one or several,
for example by making only one of the two dotted-arrow identifications in
Fig.~6, yielding black hole configurations with internal topological
structure.\footnote{The interesting case when there is only one asymptotic
region and internal structure will be the subject of a separate publication
\cite{I}.}

\bigskip
{\large III. Time Development: Enter the Raychaudhuri Equation}

\medskip
The Multi-Black-Hole initial values we discussed in the last section differ
from adS initial values only in the global structure; in particular, they
are locally homogeneous and isotropic. Therefore the time development is
completely described by that of the volume element, $\sqrt{g}$. It is
the {\em Raychaudhuri Equation} that gives us this time development.

In the context of an analysis of development in proper time $\tau$
a convenient form of the (2+1)D Raychaudhuri equation is
$$ \dot K+ K^{ij}K_{ij} + 2(p - \Lambda) = 0.$$
The local isotropy demands isotropic time development,
$K_{ij} = {1\over 2}Kg_{ij}$, with $K= {1\over\sqrt{g}}{\partial\sqrt{g}\over
\partial \tau}$.
In the absence of matter we have $p = 0$, hence
$$ \dot K + {1 \over 2} K^2 - 2\Lambda = 0$$
The solution that satisfies the initial condition $K=0$ is
$$ K = - \left({2\over\ell}\right)\tan\left({\tau\over\ell}\right)$$
Thus after a time $\tau = \pi\ell/2$, $K$ diverges. This is the first 
consequence
usually obtained from the Raychaudhuri equation: there is a conjugate point
where the normal geodesics to the initial surface intersect. Because of the
homogeneity of our model, all these geodesics intersect in one point ($r=0=t$ 
in BTZ coordinates), whether they start inside the horizon or outside.

The identification that is used to construct the initial state must of course
extend to these geodesics. It yields a nonsingular space if the corresponding
isometries are transitive, that is, if there is no fixed point.
This is of course no longer true at the point where the geodesics intersect.
In Riemannian spaces a fixed point leads to a conical singularity; in
Lorentzian spaces a fixed point in whose neighborhood the isometry is a
boost leads to a ``non-Hausdorff" singularity \cite{M}. It is customary
to regard such singularities as analogous to the black hole curvature
singularity of higher dimensions, and define a horizon that hides this
singularity in an analogous way.

The Raychaudhuri analysis specifies, in this isotropic case, the entire time
development in Raychaudhuri time\footnote{I use this convenient term
for the time-orthogonal slicing of spacetime with proper time ($\tau$) 
coordinate that corresponds to constant lapse function 
$N=|\partial/\partial\tau|$.} of any Multi-Black-Hole solution as it follows 
from time symmetric initial data. Namely, the conformal 
representation of constant-$\tau$ surfaces is the same as the initial surface 
$\tau = 0$, as for example in Fig.~4a; only the conformal factor decreases,
reaching zero at $\tau = \pi\ell/2$. But note that a BTZ $r$-coordinate, 
as in Fig.~2a, measures the circumference of circles at all times, hence
if all space sections are represented by one constant diagram in the
Poincar\'e disk, then any $r$-coordinate lines should move outward as
$\tau$ increases, to compensate for the decrease in the conformal factor.
The geodesics of test particle with zero initial velocity are represented
by a point at rest in the diagram, their inward fall being exhibited by the
outward motion of the $r =$ const curves.

This representation shows only the initial data's domain of dependence and
not its analytic continuation beyond the Cauchy horizon. For example, the
outward-moving $r=$ const lines (which have finite acceleration), as well
as geodesics with initial velocity, and light rays accumulate at the limit 
circle in a finite time. (It is amusing that their subsequent development
can be described as motion outside the limit circle, in a metric with
a changed signature.) The Raychaudhuri time slicing therefore represents
accurately only the interior part of a multi-black-hole solution. For
the closed universes of Section IIC, which have no asymptotic region, it
does represent the nonsingular part of the time development in its entirety.
The exterior of any horizon is of course the same as that of a BTZ black
hole of the same mass parameter. To understand how the exterior and interiors
fit together, so that horizons can be analyzed, we return to the stereographic
projection.

\bigskip
{\large IV. Time Development in Stereographic Projection}

\medskip

To understand the stereographic, conformal map (3b) away from the surface 
$T=0$, consider first the section $Y=0$ (Figure 7). In this case the 
projection fills the space between two limit {\em hyperbolas}. 
This is the spacetime analog of the Poincar\'e disk. We call it the 
{\em Minkowski disk}. We can again think of it as a magnifying glass 
that distorts except at the center --- away from the center the magnification 
decreases to zero in spacelike directions, and increases to infinity in 
timelike directions. Unlike the projection of the constant
curvature spaces of positive definite metric, this is not a map of the
complete spacetime, but only of the past (resp.\ future) of the ``point at 
infinity."  One can verify that the geometrical properties of the Poincar\'e 
disk still hold if circles are everywhere replaced by hyperbolas. 
For example, spacelike geodesics are hyperbolas that meet 
the limit hyperbolas normally in the Minkowski sense (or not at all).

\medskip

In Figure 7 the initial ``surface" $T=0$ that extends to infinity is projected 
by conformal distortion to a finite-length diameter of the limit hyperbola.
The $r=0$ singularities of a BTZ black hole occur 90$^o$ 
away from the initial surface, in the plane $U=0$. One of these is shown 
in the Figure as a dotted hyperbola. The projection of that hyperbola on the 
plane is also a hyperbola (not shown), asymptotic to the dotted lightlike
lines. Thus the stereoscopic projection of that section is very similar
to the Kruskal diagram of a Schwarzschild black hole, except that null
infinity $\scri$ is represented at a finite distance --- it is the 
limit hyperbola.

If we rotate the hyperboloid with initial ``surface" and singular hyperbola 
by 90$^o$ (but leave the projection center and the plane fixed), 
we thereby center our lens on the singularity, which in the
projection is now represented by the finite-length diameter of the limit
hyperbola. The initial surface is now the dotted hyperbola. Because it
lies in a plane parallel to the projection plane, the conformal factor
is constant, so that its projection involves no distortion.

\bigskip
{\large A. The BTZ black hole spacetime}

\medskip
We are now ready to consider the projection in three dimensions. The 2+1
adS space is now represented by the interior of a limit hyperboloid.
We call this interior the Minkowski ball. The horizontal section through
the center of this figure is a disk --- it is the Poincar\'e disk that
we discussed as the $T=0$ initial state. To construct black holes we now
merely have to extend the identifications introduced in Section II 
consistently throughout the Minkowski ball.

Figure 8a shows the Minkowski ball, the Poincar\'e disk embedded as an
initial surface, a geodesic through the center of the Poincar\'e disk,
and the Minkowski disk normal to the Poincar\'e disk at this geodesic.
We can move the geodesic and its Minkowski disk to another (ultraparallel)
position by means of an isometry (a kind of transvection of the Minkowski
ball, which however has fixed points -- as we shall see). The result is
shown in Figure 8b. Not surprisingly the totally geodesic Minkowski disk 
now becomes a hyperboloid. The two ultraparallels on the Poincar\'e disk 
(of Figs.~8a and b, respectively) are suitable for identification to form 
the initial state of a BTZ black hole, as we have seen in Section II. 
The identification can be extended to the corresponding Minkowski disks, 
since their intrinsic geometry is the same and their extrinsic curvature
vanishes. The resulting spacetime, that is adS space modulo the isometry,
is the unique development of the initial data -- it is a BTZ black hole
spacetime.

\medskip

The Minkowski disks of Figure 8 intersect in a spacelike hyperbola. Its
points are fixed points of the isometry, and represent the non-Hausdorff
singularity at $r=0$ of the BTZ spacetime. Figure 9 is a combination of 
Figs.~8a and b, cut up and separated along a vertical plane. Some BTZ
coordinate lines are plotted in this vertical plane, showing again their
similarity to the Kruskal coordinates of Schwarzschild spacetime. The BTZ
spacetime is the heavily-outlined part of the right half of the figure;
its front and back surfaces correspond to $\phi = 0$ and $2\pi$, respectively,
and are to be identified.

\medskip

So far our stereographic projection has been centered on the initial surface.
It is also interesting to center it on the singularity. The figure is then
a three-dimensional version of the projection discussed in connection with
Figure 7, with the dotted hyperbola being the initial surface. As before,
the singularity is then a line, and it is 
still true that on the initial surface, the conformal factor is constant,
so the initial surface can be shown without distortion, as a hyperboloid
in Minkowski space (Fig.~10). Furthermore, the plane Minkowski disks
($\phi_{BTZ}=$ const) are represented as planes intersecting the singular
line. Since that line, and hence the planes, pass through the origin,
the planes intersect the initial surface in (ultraparallel) geodesics, as
they should in order to generate a BTZ black hole after identification.
Null infinity $\scri$ now ends at a finite point, so it is easy to draw
the associated horizon --- it is just the backward Minkowski lightcone from 
that point.

\bigskip

{\large B. Multi-Black-Hole Spacetimes}

\medskip
The above spacetime construction can be extended to the multi-black-hole
case. Mathematically, the isometries of the initial surface that define
the multi-black-hole initial values extend to all of adS space,
so the multi-black-hole spacetime is the quotient of adS space by the
isometries so extended. In the stereographic projection we would have to add
further hyperboloids to Figs.~8 and 9, and double in the way described
in Section II the spacetime between them. Pairs of hyperboloids intersect
in spacelike hyperbolas, and all of the hyperboloids intersect in one
common point, where any geodesic motion with initially vanishing velocity ends 
--- this was one of the lessons learned from the Raychaudhuri
equation. Thus the non-Hausdorff singularity is represented by several
hyperbolas radiating out from the common point. Geodesics that start
out with nonzero velocity end on one of the hyperbolas, depending into
which black hole they fall. Null geodesics either end similarly or go
to a $\scri$ in an asymptotic region. The two possible behaviors allow the 
usual definition of horizons as the boundary of the region that is in the
past of $\scri$.

The identification surfaces are simplest in the stereographic projection that
is centered on the common point of the singularity, because they are then
planes as in Fig.~10. These planes form a pyramid with spacelike edges, and
the physical region is the intersection of the inside of the pyramid with
the Minkowski ball (Fig.~11). The ends of each $\scri$ occur where one of
the edges pierces the limit hyperboloid. The backward lightcones from these
ends form the horizon, as will be discussed in a separate paper \cite{I}.

\medskip

Here we will explore only the section of a four-black-hole horizon in a 
plane of symmetry S. We return to the Kruskal-like representation of a 
projection centered on the initial surface and show the relevant features 
in Figure 12. The plane of symmetry cuts through two of the horizons on 
the initial surface, labeled $h_1$ and $h_3$. The spacetime diagram shows
the entire history in S of these horizons. Unlike the single BTZ black hole
horizons, these horizons have a past endpoint on the singularity associated
with the {\it other} black hole. Thus for an observer in an external region
we have the following situation: the domain of dependence of the initial
region exterior to the horizon is the same as that of a single BTZ black hole.
However, prior to the past Cauchy horizon the geometry is different from
that of a single black hole. The observer in one exterior region can see
events from the the other black hole region, for example the other
``white hole" singularity. By looking at the past, an external observer
can tell that she is in a multi-black-hole spacetime, although no
experiments she can do in the future will reveal this.

\bigskip

{\large V. Black Holes with Angular Momentum} 

\medskip

The general BTZ black hole has angular momentum. It therefore has no global
totally geodesic spacelike surfaces to simplify an initial value description.
We therefore treat it from the spacetime point of view. Because adS space
is the universal cover, black holes with angular momentum can still be
considered a piece of adS space, suitable identified. In fact, one can
use the same piece, and the same identification surfaces, as those shown in
Figures 9 and 10. The only difference is that the isometry that defines the
identification is not simply a ``boost" about a geodesic (which would become
the non-Hausdorff singularity), but in addition it involves a boost within 
an identification surface. For example, in Figure 9 it could be a boost
in the Minkowski disk of the figure's left half, with fixed point at the
origin --- that is, a displacement of the BTZ time-coordinate in that plane.
This boost leaves the geodesic labeled $r=0$ invariant, but moves the points
within that geodesic, so that the geodesic is no longer a set of fixed points,
and hence no singularity occurs there. (With this ``boosted" identification,
this geodesic also no longer has BTZ coordinate $r=0$, because $r$ measures
the circumference of the circle at constant $r$, and that is now the finite
amount by which points move on that geodesic.)

Although the boosted identification has no fixed points, there is still a
line with $r_{BTZ}=0$, namely the one that connects identified points
and that is {\it lightlike}. (This would occur somewhere in the region that 
is spacelike related to the geodesic {\it labeled} $r=0$ in Fig.~9). Beyond
that line, timelike curves connect identified points. Therefore the line
$r_{BTZ}=0$ is still considered a kind of singularity, since it bounds
a region of closed timelike curves.

Can we make a similar boosted identification in the multi-black-hole case?
The condition is that the spacetime should look like a single BTZ black
hole with angular momentum in each asymptotic region. A possible problem
is that a boost about one horizon line is a timelike motion at the
neighboring horizon lines, but there should be no net timelike motion as
one goes once around all the black holes and comes back to the starting
point --- else we would get closed timelike lines everywhere. Therefore
not all the angular momenta can be chosen arbitrarily. It can however be
shown \cite{B} that, with a sufficient number of black holes, multi-black-hole
configurations with angular momenta are possible.

\bigskip

{\large VI. Analogous 3+1 Dimensional Black Holes}

\medskip

The BTZ black holes can be generalized in various ways to 3+1 dimensional 
gravity with negative cosmological constant. Thus one can obtain black
hole configurations of various topologies and horizon structures. For
example, the following metrics are solutions of Einstein's equations \cite{L}:
$$ds^2 = -F dt^2 + {dr^2\over F} + r^2(d\theta^2 + \theta^2 d\phi^2)
\quad {\rm with} \quad F = -{2m\over r}+{r^2\over\ell^2} \eqno(8)$$
$$ds^2 = -F dt^2 + {dr^2\over F} + r^2(d\theta^2 + \sinh^2\theta\, d\phi^2)
\quad {\rm with} \quad F = -1-{2m\over r}+{r^2\over\ell^2}.\eqno(9)$$
In a sense these interpolate between BTZ-type metrics of constant curvature,
and Schwarzschild-type metrics, with curvature increasing with decreasing $r$.
If we annul the Schwarzschild mass parameter $m$ we obtain metrics for
3+1 adS space written in BTZ-type coordinates.
The first metric is an analog of a BTZ metric with $M=0$, and the second 
metric is analogous to one with $M=1$. (These can be obtained from the
well-known Schwarzschild-deSitter metric by the substitution $\Lambda
= 1/\ell^2$, $\quad \theta~\rightarrow~i\theta, \quad r~\rightarrow~ir, \quad 
t~\rightarrow~it, \quad m~\rightarrow~im.$) 

In order that these metrics describe black holes even when $m=0$ we need
again to make identifications. In Eq (8) the surfaces $r=$ const, $t=$
const are flat, with $\theta,\, \phi$ acting as polar coordinates. Such
flat surfaces (with constant extrinsic curvature) are known in the Poincar\'e
ball as {\it horospheres} --- they are spheres that touch the limit sphere
at one point, the point at infinity of the flat surface. We can change
these to finite surfaces by means of a torus identification; that is, we
identify $x=\theta\cos\phi$ and $y=\theta\sin\phi$ periodically. The result
is an ``extremal" BTZ-type black hole, with an infinitely long, toroidal 
throat.

For a black hole with a finite throat we use the metric of Eq (9) and note
that the surfaces $r=$ const, $t=$ const have negative intrinsic curvature. 
As we found out above (for example, fig.~6), by identifications such spaces
can be given finite area (and topology with negative Euler characteristic).
This area has a finite minimum on the initial surface at the horizon, where
$F=0$.

The horizons of these black holes have various, non-spherical topologies.
This does not violate any horizon theorems, because infinity ($\scri$)
has the same unusual topology. For further examples of non-standard,
asymptotically anti-de Sitter, black holes see \cite{IS}. Multi-black-hole
geometries can similarly be constructed \cite{L}.

\bigskip

{\large VII. Conclusions}

\medskip

We have seen a remarkable variety of black hole configurations that are
possible in spaces of negative cosmological constant, even when the
gravitational field does not have local degrees of freedom. The exteriors
of such configurations have a timelike Killing vector and are therefore
static or stationary. The interior has nontrivial time development, which
can largely be derived from the Raychaudhuri equation. Because even
complicated examples can be constructed rather explicitly it is likely
that these solutions will be useful testing grounds for new ideas about
black hole physics.

\pagebreak

\small
\unitlength 0.80mm
\linethickness{0.4pt}
\begin{picture}(144.00,110.48)(-13,0)
\thicklines
\bezier{436}(43.00,20.00)(80.00,60.00)(117.00,20.00)
\bezier{160}(43.00,19.00)(42.67,14.96)(60.00,12.17)
\bezier{188}(60.00,12.17)(80.00,9.52)(100.00,12.17)
\bezier{160}(100.00,12.17)(115.67,14.08)(117.00,19.00)
\bezier{50}(43.00,19.00)(42.67,23.04)(60.00,25.83)
\bezier{60}(60.00,25.83)(80.00,28.48)(100.00,25.83)
\bezier{50}(100.00,25.83)(115.67,23.92)(117.00,19.00)
\thinlines
\put(108.00,60.00){\makebox(0,0)[cc]{X}}
\put(80.00,50.00){\vector(0,-1){10.00}}
\put(80.00,50.00){\vector(0,1){10.00}}
\put(82.00,53.00){\makebox(0,0)[cc]{$\ell$}}
\bezier{100}(43.00,100.00)(80.00,60.00)(117.00,100.00)
\thicklines
\bezier{160}(41.15,40.00)(40.80,35.76)(59.00,32.83)
\bezier{188}(59.00,32.83)(80.00,30.05)(101.00,32.83)
\bezier{160}(101.00,32.83)(117.45,34.83)(118.85,40.00)
\bezier{160}(41.15,40.00)(40.80,44.24)(59.00,47.17)
\bezier{188}(59.00,47.17)(80.00,49.95)(101.00,47.17)
\bezier{160}(101.00,47.17)(117.45,45.17)(118.85,40.00)
\put(16.00,30.00){\line(1,0){109.00}}
\thinlines
\put(40.00,50.00){\line(-6,-5){24.00}}
\put(125.00,30.00){\line(2,3){13.33}}
\put(138.33,50.00){\line(-1,0){98.7}}
\put(87.00,69.00){\makebox(0,0)[cc]{Y}}
\put(80.00,95.00){\makebox(0,0)[cc]{U}}
\bezier{70}(40.00,20.00)(60.00,40.00)(80.00,60.00)
\bezier{70}(80.00,60.00)(98.00,42.00)(120.00,20.00)
\thicklines
\put(80.00,60.00){\vector(4,3){9.00}}
\put(80.00,60.00){\vector(1,0){25.00}}
\put(80.00,60.00){\vector(0,1){30.00}}
\thinlines
\bezier{40}(43.00,101.00)(42.67,96.96)(60.00,94.17)
\bezier{45}(60.00,94.17)(80.00,91.52)(100.00,94.17)
\bezier{40}(100.00,94.17)(115.67,96.08)(117.00,101.00)
\bezier{40}(43.00,101.00)(42.67,105.04)(60.00,107.83)
\bezier{45}(60.00,107.83)(80.00,110.48)(100.00,107.83)
\bezier{40}(100.00,107.83)(115.67,105.92)(117.00,101.00)
\put(80.00,80.00){\line(1,-1){37.00}}
\put(127.00,33.00){\vector(1,-1){13.00}}
\bezier{27}(118.00,42.00)(121.67,38.33)(127.00,33.00)
\put(80.00,80.00){\line(-1,-1){37.00}}
\bezier{31}(41.00,41.00)(36.67,36.33)(30.00,30.00)
\put(30.00,30.00){\vector(-1,-1){10.00}}
\put(80.00,80.00){\line(-1,-6){9.83}}
\put(70.20,21.00){\circle*{1.00}}
\put(72.50,35.00){\circle*{1.00}}
\put(122.00,6.00){\vector(-1,1){9.00}}
\put(124.00,6.00){\makebox(0,0)[lc]{hyperboloid}}
\put(142.00,40.00){\vector(-1,0){10.00}}
\put(144.00,40.00){\makebox(0,0)[lc]{plane}}
\put(12.00,18.00){\makebox(0,0)[ct]{meets hyperboloid at $\infty$}}
\put(28.00,46.00){\makebox(0,0)[rc]{limit circle}}
\put(125.00,108.00){\vector(-4,-3){8.00}}
\put(126.00,110.00){\makebox(0,0)[lc]{other sheet of hyperboloid}}
\put(126.00,107.00){\makebox(0,0)[lt]{(not used in construction)}}
\put(73.00,21.00){\makebox(0,0)[cc]{P}}
\put(75.50,35.00){\makebox(0,0)[cc]{Q}}
\put(29.00,46.00){\vector(1,0){23.00}}
\end{picture}

\noindent
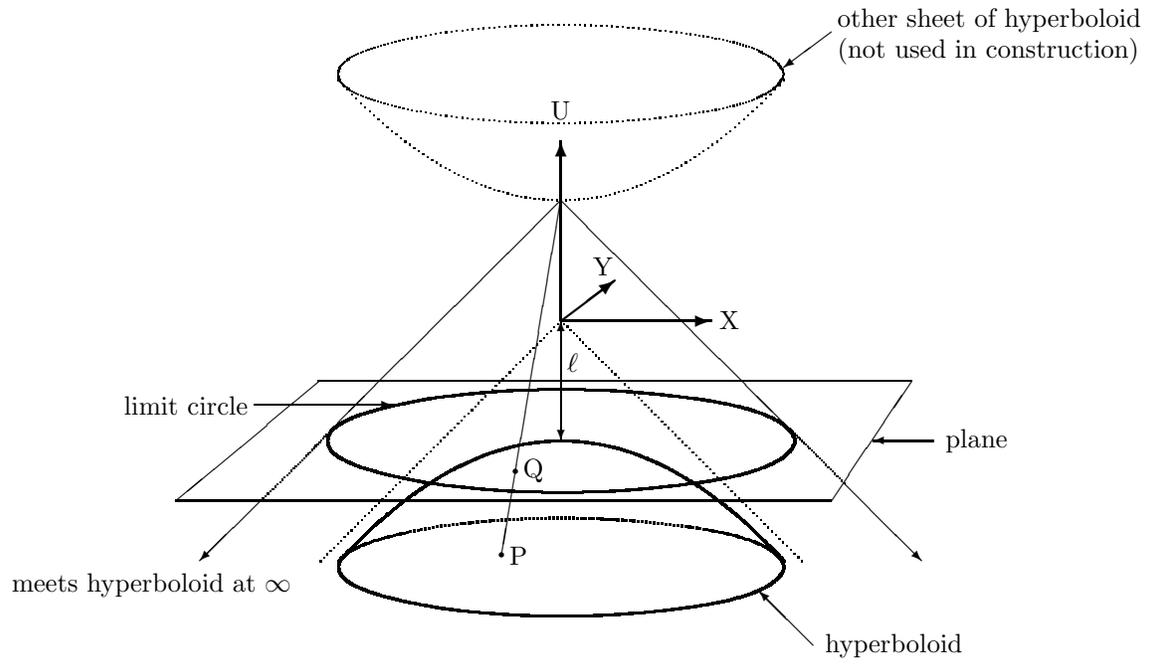
Figure 1: Stereographic projection of a spacelike hyperboloid of constant 
curvature in Minkowski space on a plane. Point P in the hyperboloid
projects to point Q in the plane. The hyperboloid is the intersection of
2+1 adS space, Eq (2), with the plane $T=0$.

\pagebreak

\unitlength 1.00mm
\linethickness{0.4pt}
\begin{picture}(133.31,49.63)
\bezier{30}(12.55,30.00)(12.55,39.74)(21.27,45.27)
\bezier{30}(21.27,45.27)(30.00,49.63)(38.73,45.27)
\bezier{30}(38.73,45.27)(47.45,39.74)(47.45,30.00)
\bezier{30}(12.55,30.00)(12.55,20.26)(21.27,14.73)
\bezier{30}(21.27,14.73)(30.00,10.37)(38.73,14.73)
\bezier{30}(38.73,14.73)(47.45,20.26)(47.45,30.00)
\thicklines
\bezier{56}(17.83,42.33)(23.17,37.50)(30.00,37.50)
\bezier{56}(42.17,42.33)(36.83,37.50)(30.00,37.50)
\bezier{56}(17.83,17.67)(23.17,22.50)(30.00,22.50)
\bezier{56}(42.17,17.67)(36.83,22.50)(30.00,22.50)
\thinlines
\put(13.00,30.00){\line(1,0){34.00}}
\bezier{144}(13.83,36.67)(30.00,29.33)(46.17,36.67)
\bezier{144}(13.83,23.33)(30.00,30.67)(46.17,23.33)
\bezier{36}(24.33,46.50)(25.50,42.00)(30.00,42.17)
\bezier{36}(35.67,46.50)(34.50,42.00)(30.00,42.17)
\bezier{36}(24.33,13.50)(25.50,18.00)(30.00,17.83)
\bezier{36}(35.67,13.50)(34.50,18.00)(30.00,17.83)
\put(30.00,47.00){\line(0,-1){34.00}}
\bezier{96}(30.00,47.50)(19.83,42.83)(19.00,30.00)
\bezier{96}(30.00,12.50)(19.83,17.17)(19.00,30.00)
\bezier{168}(30.00,47.33)(18.33,29.83)(30.00,12.83)
\bezier{96}(30.00,47.50)(40.17,42.83)(41.00,30.00)
\bezier{96}(30.00,12.50)(40.17,17.17)(41.00,30.00)
\bezier{168}(30.00,47.33)(41.67,29.83)(30.00,12.83)
\put(55.00,41.00){\vector(-2,-1){7.00}}
\put(63.50,41.00){\makebox(0,0)[cc]{$\phi =$ const}}
\put(55.00,27.00){\vector(-1,0){14.00}}
\put(63.50,27.00){\makebox(0,0)[cc]{$r=$ const}}
\bezier{64}(94.74,13.38)(90.50,13.38)(88.08,21.69)
\bezier{70}(88.08,21.69)(86.18,30.00)(88.08,38.31)
\bezier{64}(88.08,38.31)(90.50,46.62)(94.74,46.62)
\bezier{64}(124.74,13.38)(128.99,13.38)(131.41,21.69)
\bezier{70}(131.41,21.69)(133.31,30.00)(131.41,38.31)
\bezier{64}(131.41,38.31)(128.99,46.62)(124.74,46.62)
\bezier{64}(94.74,13.38)(98.99,13.38)(101.41,21.69)
\bezier{70}(101.41,21.69)(103.31,30.00)(101.41,38.31)
\bezier{64}(101.41,38.31)(98.99,46.62)(94.74,46.62)
\bezier{144}(99.00,44.00)(111.67,30.00)(122.00,44.00)
\bezier{20}(122.00,44.00)(123.5,46.33)(124.74,46.62)
\bezier{144}(99.00,16.00)(111.67,30.00)(122.00,16.00)
\bezier{20}(122.00,16.00)(123.5,13.67)(124.74,13.38)
\bezier{64}(109.89,22.75)(111.74,22.75)(112.80,26.38)
\bezier{60}(112.80,26.38)(113.63,30.00)(112.80,33.62)
\bezier{64}(112.80,33.62)(111.74,37.25)(109.89,37.25)
\bezier{132}(102.00,35.00)(109.00,29.00)(132.00,35.00)
\bezier{128}(102.00,23.00)(114.00,29.00)(130.00,19.00)
\bezier{64}(89.00,20.00)(90.67,24.33)(102.00,27.00)
\bezier{64}(88.00,38.00)(91.00,34.33)(102.00,33.00)
\put(111.00,47.00){\vector(0,-1){9.00}}
\put(111.00,48.00){\makebox(0,0)[cb]{horizon}}
\put(71.00,41.00){\vector(4,-1){20.00}}
\put(71.00,27.00){\vector(1,0){16.00}}
\put(30.00,5.00){\makebox(0,0)[cc]{(a)}}
\put(111.00,5.00){\makebox(0,0)[cc]{(b)}}
\end{picture}

\medskip \noindent
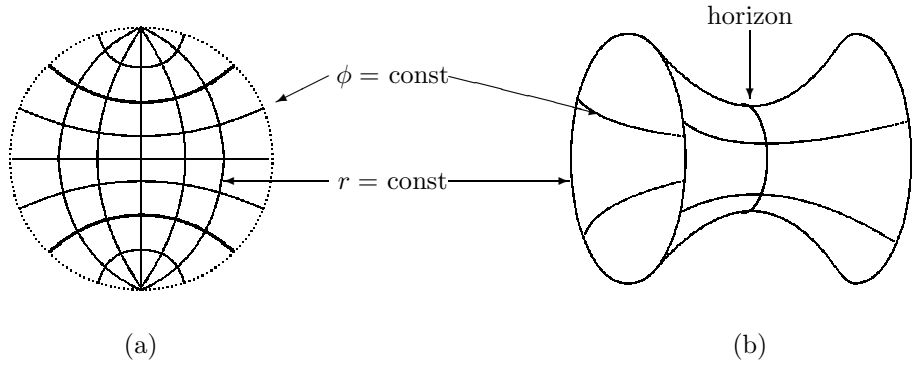
Figure 2: a. BTZ coordinate lines $r,\,\phi$ on the Poincar\'e disk, centered
at 
the horizon. All coordinate lines have constant curvature and therefore
appear as (parts of) circles. The lines $\phi =$ const have zero curvature
(geodesics) and are mutually ultraparallel. In the BTZ black hole the lines
$\phi=0$ and $\phi=2\pi$ (drawn thick) are identified. b. Sketch of part
of the surface obtained by identification. The asymptotic region at large $r$
cannot be embedded into flat space.

\pagebreak

\unitlength 1.00mm
\linethickness{0.4pt}
\begin{picture}(79.85,55.99)
\bezier{34}(20.15,23.00)(20.15,28.98)(35.08,32.38)
\bezier{30}(35.08,32.38)(50.00,35.05)(64.92,32.38)
\bezier{34}(64.92,32.38)(79.85,28.98)(79.85,23.00)
\bezier{34}(20.15,23.00)(20.15,17.02)(35.08,13.62)
\bezier{30}(35.08,13.62)(50.00,10.95)(64.92,13.62)
\bezier{34}(64.92,13.62)(79.85,17.02)(79.85,23.00)
\thicklines
\bezier{104}(21.40,19.73)(33.51,21.17)(40.00,25.14)
\bezier{104}(40.00,25.14)(46.49,28.79)(44.04,33.71)
\bezier{104}(78.60,26.27)(66.49,24.83)(60.00,20.86)
\bezier{104}(60.00,20.86)(53.51,17.21)(55.96,12.29)
\thinlines
\bezier{34}(20.15,47.00)(20.15,51.46)(35.08,54.00)
\bezier{30}(35.08,54.00)(50.00,55.99)(64.92,54.00)
\bezier{34}(64.92,54.00)(79.85,51.46)(79.85,47.00)
\bezier{34}(20.15,47.00)(20.15,42.54)(35.08,40.00)
\bezier{30}(35.08,40.00)(50.00,38.01)(64.92,40.00)
\bezier{34}(64.92,40.00)(79.85,42.54)(79.85,47.00)
\thicklines
\bezier{104}(21.40,44.56)(33.51,45.64)(40.00,48.60)
\bezier{104}(40.00,48.60)(46.49,51.32)(44.04,54.99)
\bezier{104}(78.60,49.44)(66.49,48.36)(60.00,45.40)
\bezier{104}(60.00,45.40)(53.51,42.68)(55.96,39.01)
\thinlines
\put(56.00,30.00){\vector(0,1){9.00}}
\put(56.00,30.00){\vector(0,-1){17.00}}
\put(21.00,35.00){\vector(0,1){9.00}}
\put(79.00,38.00){\vector(0,1){11.00}}
\put(79.00,38.00){\vector(0,-1){11.00}}
\put(21.00,35.00){\vector(0,-1){15.00}}
\put(44.00,39.00){\vector(0,-1){5.00}}
\put(44.00,54.00){\vector(0,1){0.2}}
\bezier{15}(44.00,39.00)(44.00,47.5)(44.00,54.00)
\put(63.00,39.00){\vector(0,-1){16.00}}
\put(63.00,46.00){\vector(0,1){0.2}}
\bezier{7}(63.00,39.00)(63.00,43.5)(63.00,46.00)
\put(39.00,39.00){\vector(0,-1){14.00}}
\put(39.00,48.00){\vector(0,1){0.2}}
\bezier{9}(39.00,39.00)(39.00,44.5)(39.00,48.00)
\end{picture}

\noindent
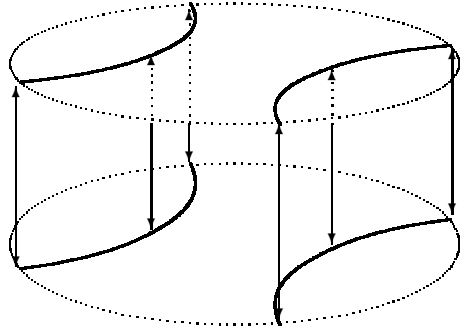
Figure 3: Constructing the initial state of a BTZ black hole by doubling
a strip of the Poincar\'e disk bounded by two ultraparallel geodesics.

\pagebreak

\unitlength 1.00mm
\linethickness{0.4pt}
\begin{picture}(152.00,58.40)(10,0)
\thicklines
\bezier{144}(24.50,42.50)(37.17,30.00)(24.50,17.50)
\bezier{140}(37.00,49.75)(41.50,32.67)(58.75,36.83)
\bezier{140}(37.00,10.33)(41.50,27.33)(58.75,23.00)
\thinlines
\bezier{64}(24.50,42.50)(29.00,48.50)(37.00,49.75)
\bezier{64}(58.75,37.17)(61.33,30.00)(58.75,23.00)
\bezier{64}(24.50,17.50)(29.00,11.50)(37.00,10.33)
\bezier{40}(37.00,49.75)(53.17,50.83)(58.75,37.17)
\bezier{40}(24.50,42.50)(15.00,30.00)(24.50,17.50)
\bezier{40}(37.00,10.33)(53.17,9.17)(58.75,23.00)
\bezier{15}(53.20,45.33)(49.60,41.87)(47.73,36.67)
\bezier{70}(47.73,36.67)(45.87,30.00)(47.87,23.33)
\bezier{15}(47.87,23.33)(49.47,18.00)(53.20,14.67)
\thicklines
\bezier{144}(100.91,30.01)(114.88,24.00)(100.91,17.99)
\bezier{140}(114.69,33.50)(119.65,25.28)(138.67,27.29)
\bezier{140}(114.69,14.54)(119.65,22.72)(138.67,20.63)
\thinlines
\bezier{64}(100.91,30.01)(105.87,32.90)(114.69,33.50)
\bezier{64}(138.67,27.45)(141.52,24.00)(138.67,20.63)
\bezier{64}(100.91,17.99)(105.87,15.10)(114.69,14.54)
\bezier{40}(114.69,33.50)(132.52,34.02)(138.67,27.45)
\bezier{40}(100.91,30.01)(90.44,24.00)(100.91,17.99)
\bezier{40}(114.69,14.54)(132.52,13.98)(138.67,20.63)
\bezier{15}(132.55,31.37)(128.58,29.71)(126.52,27.21)
\bezier{64}(126.52,27.21)(124.47,24.00)(126.68,20.79)
\bezier{15}(126.68,20.79)(128.44,18.23)(132.55,16.63)
\put(100.00,36.00){\line(-2,-3){16.00}}
\thicklines
\put(84.00,12.00){\line(1,0){68.00}}
\thinlines
\put(152.00,12.00){\line(-2,3){16.00}}
\put(136.00,36.00){\line(-1,0){36.00}}
\thicklines
\bezier{144}(100.91,53.71)(114.88,49.00)(100.91,44.29)
\bezier{140}(114.69,56.44)(119.65,50.01)(138.67,51.57)
\bezier{140}(114.69,41.59)(119.65,47.99)(138.67,46.36)
\thinlines
\bezier{64}(100.91,53.71)(105.87,55.97)(114.69,56.44)
\bezier{64}(138.67,51.70)(141.52,49.00)(138.67,46.36)
\bezier{64}(100.91,44.29)(105.87,42.03)(114.69,41.59)
\bezier{40}(114.69,56.44)(132.52,56.85)(138.67,51.70)
\bezier{40}(100.91,53.71)(90.44,49.00)(100.91,44.29)
\bezier{40}(114.69,41.59)(132.52,41.15)(138.67,46.36)
\bezier{15}(132.55,54.78)(128.58,53.47)(126.52,51.51)
\bezier{64}(126.52,51.51)(124.47,49.00)(126.68,46.49)
\bezier{15}(126.68,46.49)(128.44,44.48)(132.55,43.22)
\multiput(100.00,58.40)(-0.12,-0.14){134}{\line(0,-1){0.14}}
\thicklines
\put(84.00,39.60){\line(1,0){68.00}}
\thinlines
\multiput(152.00,39.60)(-0.12,0.14){134}{\line(0,1){0.14}}
\put(136.00,58.40){\line(-1,0){36.00}}
\put(87.00,34.00){\vector(2,-1){21.00}}
\put(87.00,38.00){\vector(2,1){21.00}}
\put(87.00,36.00){\makebox(0,0)[cc]{identify}}
\put(40.00,1.00){\makebox(0,0)[cc]{(a)}}
\put(120.00,1.00){\makebox(0,0)[cc]{(b)}}
\put(39.00,32.00){\vector(3,-1){8.00}}
\put(39.00,33.00){\makebox(0,0)[cb]{horizon}}
\end{picture}

\medskip 
\noindent
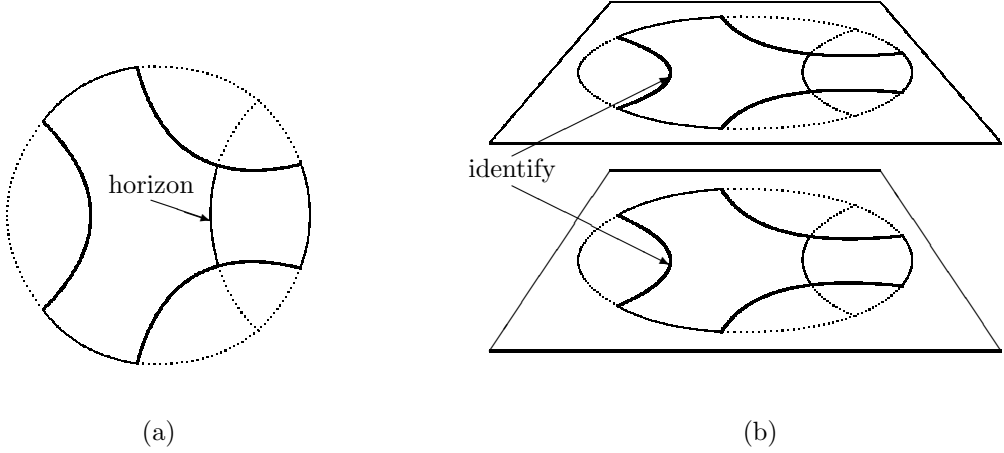
Figure 4: Construction of a three-black-hole initial geometry by doubling a 
region bounded by three geodesics in the Poincar\'e disk.

\noindent (a) Half of the initial geometry is represented by the region
bounded by the thick circular arcs. The horizon of one black hole region
(minimal geodesic between the two geodesics on the right) is shown. The
other two horizons can be obtained by 120$^o$ rotations.

\noindent (b) Two disks are placed one above the other, and the thick 
boundaries are identified vertically, as shown explicitly for the boundary 
on the left.

\pagebreak

\unitlength 0.80mm
\linethickness{0.4pt}
\begin{picture}(195.68,82.92)
(10,0)
\bezier{34}(62.00,17.08)(43.62,17.08)(33.20,33.55)
\bezier{30}(33.20,33.55)(24.99,50.00)(33.20,66.45)
\bezier{34}(33.20,66.45)(43.62,82.92)(62.00,82.92)
\bezier{33}(62.00,17.08)(80.38,17.08)(90.80,33.55)
\bezier{30}(90.80,33.55)(99.01,50.00)(90.80,66.45)
\bezier{30}(90.80,66.45)(80.38,82.92)(62.00,82.92)
\bezier{76}(47.11,51.78)(55.33,57.33)(60.22,65.33)
\bezier{56}(60.22,65.33)(66.44,62.00)(72.89,62.22)
\bezier{96}(72.89,62.22)(72.67,50.89)(80.44,41.11)
\bezier{32}(80.44,41.11)(76.67,38.00)(75.56,34.67)
\bezier{100}(75.56,34.67)(65.11,40.22)(52.00,39.33)
\bezier{56}(52.00,39.33)(51.50,46.00)(47.11,51.78)
\put(48.50,44.44){\makebox(0,0)[cc]{$h_1$}}
\put(66.67,65.00){\makebox(0,0)[cc]{$h_2$}}
\put(80.17,36.89){\makebox(0,0)[cc]{$h_3$}}
\put(53.11,59.11){\makebox(0,0)[cc]{$s_1$}}
\put(76.50,52.67){\makebox(0,0)[cc]{$s_2$}}
\put(64.00,37.00){\makebox(0,0)[cc]{$s_3$}}
\bezier{34}(158.67,17.08)(140.29,17.08)(129.87,33.55)
\bezier{30}(129.87,33.55)(121.66,50.00)(129.87,66.45)
\bezier{34}(129.87,66.45)(140.29,82.92)(158.67,82.92)
\bezier{33}(158.67,17.08)(177.05,17.08)(187.47,33.55)
\bezier{30}(187.47,33.55)(195.68,50.00)(187.47,66.45)
\bezier{30}(187.47,66.45)(177.05,82.92)(158.67,82.92)
\bezier{76}(143.78,51.78)(152.00,57.33)(156.89,65.33)
\put(145.17,44.44){\makebox(0,0)[cc]{$h_1$}}
\put(163.34,65.00){\makebox(0,0)[cc]{$h_2$}}
\put(149.78,59.11){\makebox(0,0)[cc]{$s_1$}}
\put(172.87,52.67){\makebox(0,0)[cc]{$s_2$}}
\put(160.67,37.00){\makebox(0,0)[cc]{$s_3$}}
\bezier{48}(143.78,51.78)(148.00,46.00)(148.22,40.89)
\bezier{164}(148.22,40.89)(168.22,42.67)(184.89,29.33)
\bezier{160}(184.89,29.33)(169.11,44.44)(167.78,62.22)
\bezier{48}(167.78,62.22)(161.78,61.78)(156.89,65.33)
\put(61.00,4.00){\makebox(0,0)[cc]{(a)}}
\put(160.00,4.00){\makebox(0,0)[cc]{(b)}}
\end{picture}

\medskip
\noindent
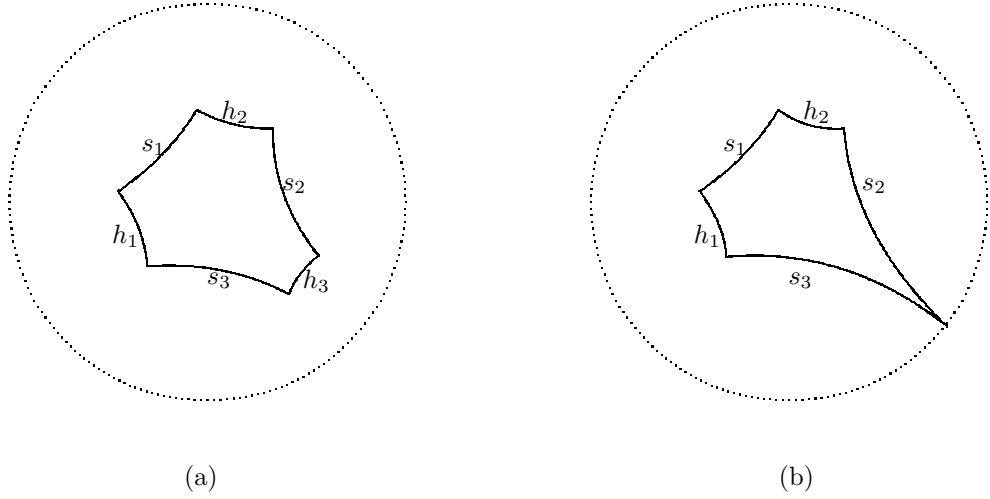
Figure 5: The parameters of multi-black-hole initial states obtained by
doubling a right-angle polygon with an even number of sides.

\noindent (a) A hexagon represents three horizons $h_i$ and three distances
$s_i$.

\noindent (b) In the limit $h_3 \rightarrow 0$ (hence $M_3 \rightarrow 0$)
the hexagon degenerates into a pentagon with one corner at infinity.

\pagebreak

\unitlength 1.10mm
\linethickness{0.4pt}
\begin{picture}(121.00,51.00)(10,5)
\bezier{13}(10.00,30.00)(10.00,36.00)(14.00,42.00)
\bezier{20}(14.00,42.00)(20.00,50.00)(30.00,50.00)
\bezier{13}(50.00,30.00)(50.00,36.00)(46.00,42.00)
\bezier{20}(46.00,42.00)(40.00,50.00)(30.00,50.00)
\bezier{13}(10.00,30.00)(10.00,24.00)(14.00,18.00)
\bezier{20}(14.00,18.00)(20.00,10.00)(30.00,10.00)
\bezier{13}(50.00,30.00)(50.00,24.00)(46.00,18.00)
\bezier{20}(46.00,18.00)(40.00,10.00)(30.00,10.00)
\bezier{6}(10.33,33.67)(13.50,33.00)(16.67,34.83)
\thicklines
\bezier{20}(16.67,34.83)(18.50,36.17)(19.83,38.67)
\thinlines
\bezier{9}(19.83,38.67)(21.67,42.83)(19.00,46.83)
\bezier{40}(14.00,41.83)(16.50,40.00)(19.83,38.67)
\thicklines
\bezier{88}(19.83,38.67)(30.00,34.67)(40.17,38.67)
\thinlines
\bezier{40}(40.17,38.67)(43.83,40.00)(46.17,42.00)
\bezier{36}(12.00,21.00)(14.83,22.17)(16.50,25.17)
\thicklines
\bezier{44}(16.50,25.17)(19.17,30.00)(16.50,34.67)
\thinlines
\bezier{36}(16.50,34.67)(15.00,37.50)(12.17,39.00)
\bezier{7}(49.67,33.67)(46.50,33.00)(43.33,34.83)
\thicklines
\bezier{20}(43.33,34.83)(41.50,36.17)(40.17,38.67)
\thinlines
\bezier{36}(48.00,21.00)(45.17,22.17)(43.50,25.17)
\thicklines
\bezier{44}(43.50,25.17)(40.83,30.00)(43.50,34.67)
\thinlines
\bezier{36}(43.50,34.67)(45.00,37.50)(47.83,39.00)
\bezier{9}(40.17,38.67)(38.33,42.83)(41.00,46.83)
\bezier{40}(14.00,18.17)(16.50,20.00)(19.83,21.33)
\thicklines
\bezier{88}(19.83,21.33)(30.00,25.33)(40.17,21.33)
\thinlines
\bezier{40}(40.17,21.33)(43.83,20.00)(46.17,18.00)
\bezier{7}(10.33,26.33)(13.50,27.00)(16.67,25.17)
\thicklines
\bezier{20}(16.67,25.17)(18.50,23.83)(19.83,21.33)
\thinlines
\bezier{9}(19.83,21.33)(21.67,17.17)(19.00,13.17)
\bezier{7}(49.67,26.33)(46.50,27.00)(43.33,25.17)
\thicklines
\bezier{20}(43.33,25.17)(41.50,23.83)(40.17,21.33)
\thinlines
\bezier{9}(40.17,21.33)(38.33,17.17)(41.00,13.17)
\thicklines
\bezier{20}(30.00,36.50)(30.00,30.00)(30.00,23.50)
\bezier{20}(71.67,27.17)(71.17,30.83)(72.83,31.00)
\bezier{36}(72.83,31.00)(75.83,29.83)(77.00,24.00)
\bezier{24}(77.00,24.00)(77.17,20.50)(74.83,21.50)
\bezier{28}(74.83,21.50)(72.50,23.67)(71.67,27.00)
\bezier{36}(80.50,36.50)(80.00,40.50)(84.00,41.17)
\bezier{32}(84.00,41.17)(87.50,42.00)(89.17,38.00)
\bezier{132}(76.50,21.50)(93.50,28.67)(101.00,16.00)
\bezier{48}(80.50,36.50)(80.67,32.67)(72.83,31.00)
\bezier{28}(98.67,42.67)(97.83,40.50)(102.50,38.33)
\bezier{32}(102.50,38.33)(107.83,36.33)(109.33,37.50)
\bezier{24}(109.33,37.50)(109.50,39.33)(105.83,41.67)
\bezier{32}(105.83,41.67)(101.00,43.83)(98.83,42.67)
\bezier{88}(89.17,38.00)(93.17,30.33)(98.50,42.50)
\bezier{28}(91.50,23.67)(89.33,24.50)(88.67,28.67)
\bezier{32}(88.67,28.67)(89.33,34.17)(92.17,35.00)
\bezier{72}(110.83,20.67)(106.83,29.00)(109.33,37.50)
\bezier{28}(101.00,16.00)(103.17,14.67)(107.33,16.50)
\bezier{24}(107.33,16.50)(110.67,18.17)(110.83,20.67)
\put(30.00,3.00){\makebox(0,0)[cc]{(a)}}
\put(92.00,3.00){\makebox(0,0)[cc]{(b)}}
\put(73.00,27.00){\vector(1,0){0.2}}
\bezier{25}(70.00,39.00)(60.67,27.00)(73.00,27.00)
\put(82.00,42.00){\vector(3,-2){0.2}}
\bezier{20}(70.00,39.00)(76.00,47.33)(82.00,42.00)
\put(105.00,40.00){\vector(-2,-1){0.2}}
\bezier{35}(121.00,30.00)(121.00,51.00)(105.00,40.00)
\put(109.00,16.00){\vector(-2,1){0.2}}
\bezier{30}(121.00,30.00)(121.00,9.00)(109.00,16.00)
\end{picture}

\vskip 0.5cm
\noindent 
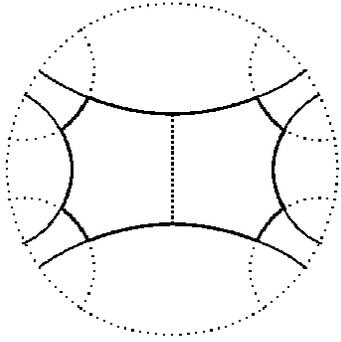
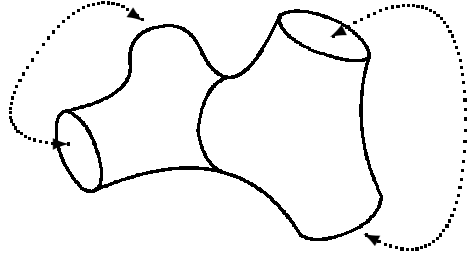
Figure 6: (a) A diagram for four black holes can be thought of as two
three-black-hole configurations joined at the thick dotted line.

\noindent (b) A 3D picture to give an idea of the result of cutting the
four-black-hole configuration at the thick dotted line, rotating, and
rejoining.

\pagebreak

\unitlength 1.00mm
\linethickness{0.4pt}
\begin{picture}(115.44,75.50)
\bezier{110}(93.37,67.00)(99.57,75.50)(105.77,67.46)
\bezier{92}(87.17,39.96)(87.17,57.46)(93.37,67.00)
\bezier{184}(105.77,67.46)(111.97,57.46)(111.97,39.96)
\bezier{92}(87.17,39.96)(87.17,22.47)(93.37,13.00)
\bezier{110}(93.37,13.00)(99.57,4.50)(105.77,12.54)
\bezier{184}(105.77,12.54)(111.97,22.47)(111.97,39.96)
\bezier{64}(63.00,53.00)(70.00,49.50)(77.00,53.00)
\bezier{150}(77.00,53.00)(85.00,57.33)(99.00,71.00)
\bezier{150}(63.00,53.00)(55.00,57.33)(41.00,71.00)
\bezier{64}(63.00,27.00)(70.00,30.50)(77.00,27.00)
\bezier{150}(77.00,27.00)(85.00,22.67)(99.00,9.00)
\bezier{150}(63.00,27.00)(55.00,22.67)(41.00,9.00)
\put(90.00,59.00){\line(4,-3){22.00}}
\bezier{184}(30.17,39.96)(30.17,57.46)(34.80,67.39)
\bezier{180}(34.80,67.39)(39.43,75.22)(44.05,67.39)
\bezier{184}(44.05,67.39)(48.68,57.46)(48.68,39.96)
\bezier{184}(30.17,39.96)(30.17,22.47)(34.80,12.54)
\bezier{180}(34.80,12.54)(39.43,4.71)(44.05,12.54)
\bezier{184}(44.05,12.54)(48.68,22.47)(48.68,39.96)
\put(90.00,59.00){\line(-1,0){58.00}}
\thicklines
\put(49.00,42.50){\line(1,0){63.00}}
\put(32.00,59.00){\line(1,-1){16.50}}
\bezier{50}(47.00,44.00)(69.00,51.00)(91.00,58.00)
\bezier{70}(33.00,58.00)(71.50,51.00)(110.00,44.00)
\bezier{140}(34.00,57.00)(56.05,51.08)(46.00,45.00)
\bezier{128}(92.00,57.00)(85.83,51.08)(108.00,45.00)
\thinlines
\put(70.00,40.00){\vector(0,1){6.00}}
\put(68.00,42.00){\vector(-1,1){0.2}}
\multiput(70.00,40.00)(-0.12,0.12){17}{\line(0,1){0.12}}
\put(70.00,40.00){\vector(1,0){8.00}}
\put(79.00,40.00){\makebox(0,0)[lc]{X}}
\put(70.00,47.00){\makebox(0,0)[cb]{U}}
\put(68.00,42.00){\makebox(0,0)[rt]{T}}
\put(70.00,28.80){\circle*{1.00}}
\put(70.00,20.00){\vector(0,1){8.00}}
\put(70.00,19.00){\makebox(0,0)[ct]{projection center}}
\put(65.00,64.00){\vector(4,-3){6.00}}
\put(63.50,64.50){\makebox(0,0)[cb]{plane}}
\put(28.00,54.00){\vector(1,0){14.00}}
\put(27.00,55.00){\makebox(0,0)[rb]{limit}}
\put(27.00,53.50){\makebox(0,0)[rt]{hyperbola}}
\bezier{25}(70.00,28.75)(67.63,28.75)(66.29,34.37)
\bezier{35}(66.29,34.37)(65.23,40.00)(66.29,45.62)
\bezier{25}(66.29,45.62)(67.63,51.25)(70.00,51.25)
\bezier{60}(70.00,28.75)(72.37,28.75)(73.71,34.37)
\bezier{80}(73.71,34.37)(74.77,40.00)(73.71,45.62)
\bezier{60}(73.71,45.62)(72.37,51.25)(70.00,51.25)
\thicklines
\bezier{50}(48.41,31.57)(74.49,40.20)(111.55,31.37)
\thinlines
\put(51.00,70.00){\vector(-1,-3){2.00}}
\put(50.96,70.39){\makebox(0,0)[cb]{$T=0$}}
\put(115.22,25.11){\vector(-4,3){9.11}}
\put(115.44,25.11){\makebox(0,0)[lc]{$r=0$}}
\put(79.00,69.00){\vector(-1,-2){8.50}}
\put(79.00,70.00){\makebox(0,0)[cb]{center of projection}}
\put(48.33,51.33){\line(1,0){44.00}}
\put(93.00,61.00){\vector(-1,-1){9.00}}
\put(93.33,61.00){\makebox(0,0)[lb]{diameter}}
\end{picture}

\noindent 
Figure 7: Stereographic projection of (1+1) adS space onto a plane of
(1+1) Minkowski space. Note the difference between what we call 
``projection center" and ``center of projection."

\pagebreak

{\small
\unitlength 1.00mm
\linethickness{0.4pt}
\begin{picture}(139.17,85)(8,0)
\bezier{64}(30.00,68.00)(30.00,72.84)(35.86,76.27)
\bezier{64}(35.86,76.27)(41.72,79.69)(50.00,79.69)
\bezier{64}(64.14,76.27)(58.28,79.69)(50.00,79.69)
\bezier{64}(64.14,76.27)(70.00,72.84)(70.00,68.00)
\bezier{64}(70.00,68.00)(70.00,63.16)(64.14,59.73)
\bezier{64}(50.00,56.31)(58.28,56.31)(64.14,59.73)
\bezier{64}(50.00,56.31)(41.72,56.31)(35.86,59.73)
\bezier{64}(39.90,45.00)(39.90,47.45)(42.86,49.18)
\bezier{64}(42.86,49.18)(45.82,50.91)(50.00,50.91)
\bezier{64}(57.14,49.18)(54.18,50.91)(50.00,50.91)
\bezier{64}(57.14,49.18)(60.10,47.45)(60.10,45.00)
\bezier{64}(60.10,45.00)(60.10,42.55)(57.14,40.82)
\bezier{64}(50.00,39.09)(54.18,39.09)(57.14,40.82)
\bezier{64}(50.00,39.09)(45.82,39.09)(42.86,40.82)
\bezier{64}(42.86,40.82)(39.90,42.55)(39.90,45.00)
\bezier{64}(30.00,68.00)(30.00,63.16)(35.86,59.73)
\bezier{64}(30.17,22.00)(30.17,26.84)(36.03,30.27)
\bezier{64}(36.03,30.27)(41.89,33.69)(50.17,33.69)
\bezier{64}(64.31,30.27)(58.45,33.69)(50.17,33.69)
\bezier{64}(64.31,30.27)(70.17,26.84)(70.17,22.00)
\bezier{64}(70.17,22.00)(70.17,17.16)(64.31,13.73)
\bezier{64}(50.17,10.31)(58.45,10.31)(64.31,13.73)
\bezier{64}(50.17,10.31)(41.89,10.31)(36.03,13.73)
\bezier{64}(30.17,22.00)(30.17,17.16)(36.03,13.73)
\multiput(42.67,49.00)(0.21,-0.12){71}{\line(1,0){0.21}}
\bezier{216}(36.00,76.33)(49.17,44.67)(35.83,30.17)
\bezier{168}(33.17,61.50)(46.50,45.00)(33.33,28.50)
\multiput(36.00,76.17)(0.20,-0.12){137}{\line(1,0){0.20}}
\multiput(64.00,13.50)(-0.20,0.12){138}{\line(-1,0){0.20}}
\bezier{220}(64.00,13.50)(50.00,44.67)(64.17,59.83)
\bezier{176}(67.83,27.67)(53.17,45.00)(66.67,61.50)
\put(50.00,68.00){\line(0,-1){46.00}}
\put(50.00,81.00){\vector(0,-1){12}}
\put(50.00,82.00){\makebox(0,0)[cb]{Minkowski disk}}
\bezier{64}(129.00,62.15)(120.72,62.15)(114.86,65.58)
\bezier{64}(129.00,44.94)(124.82,44.94)(121.86,46.67)
\bezier{64}(121.86,46.67)(118.90,48.40)(118.90,50.85)
\bezier{64}(109.00,73.85)(109.00,69.01)(114.86,65.58)
\bezier{64}(99.00,68.00)(99.00,72.84)(104.86,76.27)
\bezier{64}(104.86,76.27)(110.72,79.69)(119.00,79.69)
\bezier{64}(133.14,76.27)(127.28,79.69)(119.00,79.69)
\bezier{64}(133.14,76.27)(139.00,72.84)(139.00,68.00)
\bezier{64}(139.00,68.00)(139.00,63.16)(133.14,59.73)
\bezier{64}(119.00,56.31)(127.28,56.31)(133.14,59.73)
\bezier{64}(119.00,56.31)(110.72,56.31)(104.86,59.73)
\bezier{64}(108.90,45.00)(108.90,47.45)(111.86,49.18)
\bezier{64}(111.86,49.18)(114.82,50.91)(119.00,50.91)
\bezier{64}(129.10,45.00)(129.10,42.55)(126.14,40.82)
\bezier{64}(119.00,39.09)(123.18,39.09)(126.14,40.82)
\bezier{64}(119.00,39.09)(114.82,39.09)(111.86,40.82)
\bezier{64}(111.86,40.82)(108.90,42.55)(108.90,45.00)
\bezier{64}(99.00,68.00)(99.00,63.16)(104.86,59.73)
\bezier{36}(110.83,78.62)(108.83,76.48)(109.00,73.85)
\bezier{36}(129.00,62.06)(133.50,62.06)(137.33,63.13)
\bezier{64}(129.17,16.15)(120.89,16.15)(115.03,19.58)
\bezier{64}(109.17,27.85)(109.17,23.01)(115.03,19.58)
\bezier{64}(99.17,22.00)(99.17,26.84)(105.03,30.27)
\bezier{64}(105.03,30.27)(110.89,33.69)(119.17,33.69)
\bezier{64}(133.31,30.27)(127.45,33.69)(119.17,33.69)
\bezier{64}(133.31,30.27)(139.17,26.84)(139.17,22.00)
\bezier{64}(139.17,22.00)(139.17,17.16)(133.31,13.73)
\bezier{64}(119.17,10.31)(127.45,10.31)(133.31,13.73)
\bezier{64}(119.17,10.31)(110.89,10.31)(105.03,13.73)
\bezier{64}(99.17,22.00)(99.17,17.16)(105.03,13.73)
\bezier{36}(111.00,32.62)(109.00,30.48)(109.17,27.85)
\bezier{36}(129.17,16.06)(133.67,16.06)(137.50,17.13)
\bezier{168}(102.17,61.50)(115.50,45.00)(102.33,28.50)
\bezier{176}(136.83,27.67)(122.17,45.00)(135.67,61.50)
\bezier{220}(114.83,65.50)(129.17,50.50)(115.00,19.50)
\bezier{184}(129.00,62.00)(129.00,50.67)(129.00,16.00)
\bezier{192}(122.00,17.00)(129.00,50.67)(122.00,62.83)
\bezier{232}(110.50,23.67)(129.17,50.00)(111.33,68.17)
\bezier{188}(125.50,16.50)(129.17,50.50)(125.50,62.33)
\bezier{228}(137.17,17.00)(121.50,49.67)(137.17,63.00)
\bezier{72}(133.33,16.33)(132.33,24.17)(130.83,33.83)
\bezier{32}(133.00,62.17)(132.00,59.83)(131.17,54.17)
\bezier{204}(118.50,18.17)(128.83,50.33)(118.50,63.83)
\bezier{228}(112.67,21.33)(129.00,50.17)(112.67,67.00)
\bezier{240}(109.50,26.00)(129.17,49.33)(109.50,71.17)
\bezier{60}(110.83,78.67)(114.33,71.67)(116.67,64.67)
\bezier{40}(109.00,73.83)(112.50,70.33)(114.83,65.50)
\bezier{88}(110.83,32.67)(120.17,41.33)(119.00,50.67)
\bezier{56}(109.17,28.50)(113.67,33.50)(116.83,39.67)
\put(52.00,66.00){\line(0,-1){22.00}}
\put(52.00,44.00){\line(0,-1){5.00}}
\put(52.00,39.00){\line(0,-1){18.00}}
\put(54.00,65.00){\line(0,-1){22.00}}
\put(54.00,43.00){\line(0,-1){4.00}}
\put(54.00,39.00){\line(0,-1){19.00}}
\put(56.00,64.00){\line(0,-1){23.00}}
\put(56.00,41.00){\line(0,-1){1.00}}
\put(56.00,40.00){\line(0,-1){21.00}}
\put(58.00,63.00){\line(0,-1){14.00}}
\put(60.00,61.00){\line(0,-1){7.00}}
\put(62.00,60.00){\line(0,-1){3.00}}
\put(58.00,31.00){\line(0,-1){14.00}}
\put(60.00,24.00){\line(0,-1){8.00}}
\put(62.00,19.00){\line(0,-1){4.00}}
\put(48.00,69.00){\line(0,-1){23.00}}
\put(48.00,46.00){\line(0,-1){7.00}}
\put(48.00,39.00){\line(0,-1){16.00}}
\put(46.00,70.00){\line(0,-1){23.00}}
\put(46.00,47.00){\line(0,-1){8.00}}
\put(46.00,39.00){\line(0,-1){15.00}}
\put(44.00,71.00){\line(0,-1){23.00}}
\put(44.00,48.00){\line(0,-1){8.00}}
\put(44.00,40.00){\line(0,-1){14.00}}
\put(42.00,72.00){\line(0,-1){16.00}}
\put(42.00,40.00){\line(0,-1){13.00}}
\put(40.00,74.00){\line(0,-1){9.00}}
\put(40.00,36.00){\line(0,-1){8.00}}
\put(38.00,75.00){\line(0,-1){3.00}}
\put(38.00,33.00){\line(0,-1){4.00}}
\put(50.00,2.00){\makebox(0,0)[cc]{(a)}}
\put(120.00,2.00){\makebox(0,0)[cc]{(b)}}
\put(85.00,52.00){\makebox(0,0)[ct]{limit hyperboloid}}
\put(72.00,45.00){\vector(-1,0){11.00}}
\put(97.00,45.00){\vector(1,0){11.00}}
\put(85.00,45.00){\makebox(0,0)[cc]{Poincar\'e disk}}
\put(70.00,51.00){\vector(-3,2){6.00}}
\put(99.00,51.00){\vector(3,2){6.00}}
\end{picture}
}

\bigskip 
\noindent 
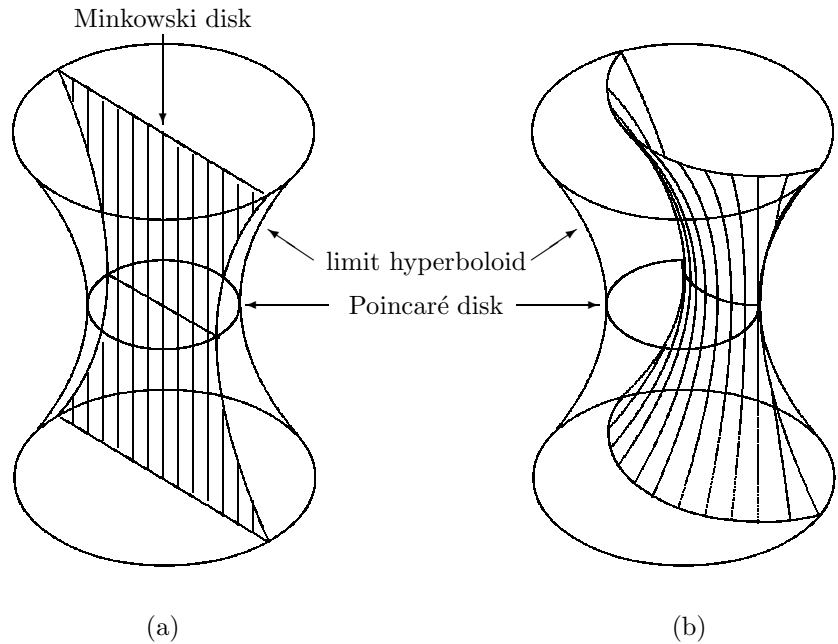
Figure 8: The Minkowski ball and two totally geodesic surfaces in it, both
normal to the same Poincar\'e disk. The totally geodesic
subspaces are striped only for identification, the lines (except the central 
one) are not geodesics.

\pagebreak

\unitlength 1.00mm
\linethickness{0.4pt}
\begin{picture}(120.17,80.54)
\thicklines
\bezier{64}(100.00,39.94)(95.82,39.94)(92.86,41.67)
\bezier{64}(92.86,41.67)(89.90,43.40)(89.90,45.85)
\bezier{32}(82.86,44.18)(85.82,45.91)(90.00,45.91)
\thinlines
\bezier{32}(97.14,44.18)(94.18,45.91)(90.00,45.91)
\bezier{32}(97.14,44.18)(100.10,42.45)(100.10,40.00)
\thicklines
\bezier{32}(100.10,40.00)(100.10,37.55)(97.14,35.82)
\multiput(82.67,44.00)(0.21,-0.12){70}{\line(1,0){0.21}}
\bezier{160}(80.00,68.83)(89.67,46.00)(100.17,57.00)
\bezier{160}(80.17,22.83)(89.83,34.00)(100.33,11.00)
\bezier{140}(86.83,57.00)(94.83,44.67)(82.00,27.70)
\bezier{216}(76.00,71.33)(89.17,39.67)(75.80,25.20)
\bezier{220}(104.00,8.50)(90.00,39.67)(104.17,54.83)
\bezier{64}(56.00,22.85)(47.72,22.85)(41.86,19.42)
\bezier{64}(36.00,11.15)(36.00,15.99)(41.86,19.42)
\bezier{64}(26.00,17.00)(26.00,12.16)(31.86,8.73)
\bezier{64}(46.00,28.69)(54.28,28.69)(60.14,25.27)
\bezier{64}(46.00,28.69)(37.72,28.69)(31.86,25.27)
\bezier{64}(35.90,40.00)(35.90,37.55)(38.86,35.82)
\bezier{64}(46.00,45.91)(50.18,45.91)(53.14,44.18)
\bezier{64}(46.00,45.91)(41.82,45.91)(38.86,44.18)
\bezier{64}(38.86,44.18)(35.90,42.45)(35.90,40.00)
\bezier{64}(26.00,17.00)(26.00,21.84)(31.86,25.27)
\bezier{64}(56.17,68.85)(47.89,68.85)(42.03,65.42)
\bezier{64}(36.17,57.15)(36.17,61.99)(42.03,65.42)
\bezier{64}(26.17,63.00)(26.17,58.16)(32.03,54.73)
\bezier{64}(46.17,74.69)(54.45,74.69)(60.31,71.27)
\bezier{64}(46.17,74.69)(37.89,74.69)(32.03,71.27)
\bezier{64}(26.17,63.00)(26.17,67.84)(32.03,71.27)
\multiput(38.67,36.00)(0.21,0.12){71}{\line(1,0){0.21}}
\bezier{216}(32.00,8.67)(45.17,40.33)(31.83,54.83)
\bezier{160}(36.00,11.17)(45.67,34.00)(56.17,23.00)
\bezier{160}(36.17,57.17)(45.83,46.00)(56.33,69.00)
\bezier{168}(29.17,23.50)(42.50,40.00)(29.33,56.50)
\bezier{220}(60.00,71.50)(46.00,40.33)(60.17,25.17)
\thinlines
\put(58.00,70.00){\line(-2,-5){24.00}}
\put(34.00,56.00){\line(3,-4){24.00}}
\bezier{220}(35.00,11.00)(46.00,42.44)(57.50,23.30)
\bezier{16}(94.89,37.00)(96.89,38.22)(96.67,40.22)
\bezier{24}(84.67,43.00)(87.22,44.67)(90.22,44.11)
\thicklines
\put(76.00,71.00){\line(2,-1){4.00}}
\put(100.00,57.00){\line(2,-1){4.00}}
\thinlines
\bezier{21}(80.00,68.85)(80.00,73.69)(85.86,77.11)
\bezier{21}(85.86,77.11)(91.72,80.54)(100.00,80.54)
\bezier{21}(114.14,77.11)(108.28,80.54)(100.00,80.54)
\bezier{21}(114.14,77.11)(120.00,73.69)(120.00,68.85)
\bezier{21}(120.00,68.85)(120.00,64.01)(114.14,60.58)
\bezier{21}(100.00,57.15)(108.28,57.15)(114.14,60.58)
\bezier{10}(89.90,45.85)(89.90,48.29)(92.86,50.02)
\bezier{10}(92.86,50.02)(95.82,51.75)(100.00,51.75)
\bezier{10}(107.14,50.02)(104.18,51.75)(100.00,51.75)
\bezier{10}(107.14,50.02)(110.10,48.29)(110.10,45.85)
\bezier{10}(110.10,45.85)(110.10,43.40)(107.14,41.67)
\bezier{10}(100.00,39.94)(104.18,39.94)(107.14,41.67)
\bezier{21}(80.17,22.85)(80.17,27.69)(86.03,31.11)
\bezier{21}(86.03,31.11)(91.89,34.54)(100.17,34.54)
\bezier{21}(114.31,31.11)(108.45,34.54)(100.17,34.54)
\bezier{21}(114.31,31.11)(120.17,27.69)(120.17,22.85)
\bezier{21}(120.17,22.85)(120.17,18.01)(114.31,14.58)
\bezier{21}(100.17,11.15)(108.45,11.15)(114.31,14.58)
\bezier{70}(117.83,28.50)(103.17,45.83)(116.67,62.33)
\bezier{84}(33.00,55.50)(41.50,43.83)(41.33,37.33)
\bezier{116}(41.33,37.33)(40.67,28.83)(33.33,9.67)
\bezier{116}(58.83,70.67)(50.67,50.83)(50.50,42.83)
\bezier{220}(57.00,69.33)(46.33,38.00)(35.17,56.50)
\bezier{12}(90.00,40.07)(91.47,40.87)(92.53,41.67)
\bezier{88}(50.50,42.83)(49.67,37.33)(59.00,24.33)
\put(69.00,51.00){\makebox(0,0)[cc]{$r=$ const}}
\put(64.00,60.00){\vector(-1,0){12.50}}
\put(74.00,60.00){\vector(1,0){10.00}}
\put(69.00,60.00){\makebox(0,0)[cc]{$r=0$}}
\put(64.00,44.00){\vector(-1,0){10.00}}
\put(74.00,44.00){\vector(1,0){8.00}}
\put(69.00,44.00){\makebox(0,0)[cc]{$t=0$}}
\put(77.00,50.50){\vector(4,-3){9.00}}
\put(61.50,51.00){\vector(-1,0){9.50}}
\bezier{88}(82.00,74.00)(93.50,76.33)(103.00,72.00)
\bezier{84}(103.00,72.00)(113.00,66.83)(108.50,58.17)
\bezier{84}(100.17,40.00)(100.50,31.50)(108.17,22.17)
\bezier{88}(82.00,28.00)(93.50,30.33)(103.00,26.00)
\bezier{84}(103.00,26.00)(113.00,20.83)(108.50,12.17)
\thicklines
\bezier{36}(100.00,57.00)(104.53,57.40)(108.53,58.20)
\bezier{20}(80.00,69.00)(80.00,71.67)(81.73,73.80)
\put(60.33,71.50){\line(-5,-3){4.17}}
\put(36.17,57.17){\line(-5,-3){4.17}}
\put(32.00,8.67){\line(5,3){4.17}}
\put(60.40,25.30){\line(-5,-3){4.20}}
\put(75.80,25.20){\line(5,-3){4.33}}
\put(104.20,8.45){\line(-5,3){3.83}}
\bezier{32}(100.17,11.00)(104.33,11.17)(108.50,12.00)
\bezier{24}(80.17,22.83)(79.67,25.00)(82.00,27.67)
\bezier{124}(108.50,12.30)(99.50,29.00)(100.00,41.00)
\bezier{80}(82.00,74.00)(86.50,63.50)(89.00,55.00)
\bezier{28}(76.00,71.00)(78.67,73.17)(82.00,74.00)
\bezier{20}(108.50,58.17)(106.67,56.17)(104.33,54.83)
\bezier{84}(100.17,40.00)(100.50,50.67)(108.33,58.00)
\bezier{28}(76.00,25.00)(78.67,27.17)(82.00,28.00)
\bezier{20}(108.50,12.17)(106.67,10.17)(104.33,8.83)
\end{picture}

\medskip
\noindent 
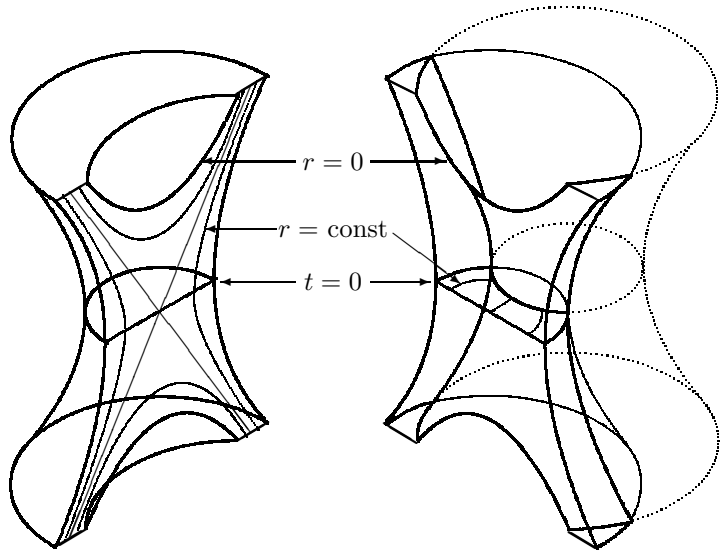
Figure 9: The two Minkowski disks of Fig.~8 in a single picture. To show
more clearly the one that is part of a hyperboloid, the rest of that
hyperboloid is drawn dotted.

\pagebreak

\unitlength 1.00mm
\linethickness{0.4pt}
\begin{picture}(83.03,56.03)
\bezier{22}(38.00,53.28)(44.83,56.03)(51.83,53.28)
\bezier{26}(51.83,53.28)(58.00,49.92)(52.17,46.79)
\bezier{26}(52.17,46.79)(45.00,43.89)(38.00,46.72)
\bezier{26}(38.00,46.72)(32.00,50.00)(38.00,53.28)
\bezier{26}(64.99,30.15)(83.03,20.79)(65.96,12.07)
\bezier{26}(65.96,12.07)(45.00,3.98)(24.52,11.85)
\put(35.50,48.33){\line(6,1){18.83}}
\bezier{280}(37.00,11.00)(46.83,44.83)(66.00,15.83)
\put(30.33,9.83){\line(6,1){42.83}}
\bezier{100}(48.83,29.00)(45.83,39.17)(43.00,39.83)
\bezier{300}(54.33,51.50)(54.00,46.17)(73.00,16.83)
\bezier{300}(35.67,48.33)(38.17,42.17)(30.17,9.83)
\bezier{65}(54.33,51.50)(51.17,45.33)(54.83,37.17)
\put(17.83,25.00){\line(6,1){6.17}}
\thicklines
\bezier{30}(44.50,12.17)(20.33,16.50)(28.17,25.33)
\bezier{20}(28.17,25.33)(35.50,33.00)(43.00,39.83)
\put(43.00,39.83){\line(1,1){11.33}}
\thinlines
\bezier{34}(55.00,50.00)(56.33,41.50)(71.67,25.50)
\bezier{38}(35.00,50.00)(35.00,42.83)(17.67,25.00)
\bezier{24}(17.67,25.00)(11.00,18.50)(25.83,11.33)
\bezier{136}(23.00,25.00)(15.67,16.00)(37.00,11.00)
\bezier{124}(37.00,11.00)(54.83,8.50)(66.00,15.83)
\bezier{48}(66.00,15.83)(71.50,20.33)(67.67,24.00)
\bezier{304}(67.67,24.00)(45.33,55.50)(23.00,25.00)
\bezier{170}(35.67,48.33)(33.50,39.00)(17.50,25.00)
\bezier{10}(64.64,30.17)(67.28,33.92)(59.50,34.19)
\bezier{10}(62.83,33.08)(64.92,36.42)(56.58,36.28)
\bezier{10}(60.89,36.69)(62.83,38.36)(54.36,38.92)
\bezier{15}(71.31,19.61)(75.33,27.11)(63.81,29.61)
\bezier{12}(67.56,25.44)(71.31,30.72)(62.42,30.86)
\bezier{7}(59.36,39.47)(60.19,40.86)(53.39,41.56)
\bezier{5}(57.42,42.67)(58.53,43.92)(53.11,44.06)
\bezier{4}(56.03,45.58)(56.86,46.14)(53.25,47.11)
\bezier{10}(66.31,27.39)(68.94,32.25)(61.17,32.11)
\bezier{7}(69.50,22.39)(72.00,26.56)(68.53,28.92)
\thicklines
\put(54.22,51.00){\line(-1,-4){5.52}}
\bezier{16}(48.53,28.92)(46.44,20.58)(44.50,12.25)
\thinlines
\put(34.64,54.75){\vector(2,-1){8.47}}
\put(34.36,54.47){\makebox(0,0)[rc]{singularity}}
\put(24.92,37.5){\vector(1,-1){5.3}}
\put(24.50,38.08){\makebox(0,0)[rc]{initial surface}}
\end{picture}

\medskip

\noindent 
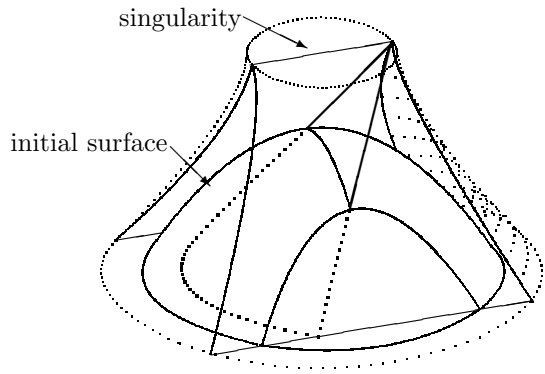
Figure 10: Stereographic projection of a BTZ black hole centered on the
non-Hausdorff singularity. The two planes that intersect in the
singularity are to be identified. One of the regions $\scri$ 
is shaded with dots, and its horizon is shown in heavy outline.
(The initial surface is drawn as non-transparent, except for the horizon. 
Therefore we do not see the hyperbolic intersection of the plane in the
back with the initial surface.)

\pagebreak

\unitlength 1.00mm
\linethickness{0.4pt}
\begin{picture}(84.04,68.00)
\bezier{30}(20.15,32.00)(20.15,38.59)(35.08,42.34)
\bezier{30}(64.92,42.34)(79.85,38.59)(79.85,32.00)
\bezier{30}(20.15,32.00)(20.15,25.41)(35.08,21.66)
\bezier{40}(35.08,21.66)(50.00,18.71)(64.92,21.66)
\bezier{30}(64.92,21.66)(79.85,25.41)(79.85,32.00)
\bezier{14}(34.17,60.00)(34.17,63.17)(42.09,64.97)
\bezier{20}(42.09,64.97)(50.00,66.39)(57.91,64.97)
\bezier{14}(57.91,64.97)(65.83,63.17)(65.83,60.00)
\bezier{14}(34.17,60.00)(34.17,56.83)(42.09,55.03)
\bezier{20}(42.09,55.03)(50.00,53.61)(57.91,55.03)
\bezier{14}(57.91,55.03)(65.83,56.83)(65.83,60.00)
\bezier{264}(24.00,35.00)(50.00,55.00)(76.00,35.00)
\bezier{56}(39.30,55.79)(46.32,39.30)(56.32,20.53)
\bezier{56}(60.70,55.79)(53.68,39.30)(43.68,20.53)
\put(50.00,31.00){\circle*{0.00}}
\put(49.00,30.00){\circle*{0.00}}
\put(50.00,29.00){\circle*{0.00}}
\put(49.00,28.00){\circle*{0.00}}
\put(50.00,27.00){\circle*{0.00}}
\put(48.00,27.00){\circle*{0.00}}
\put(49.00,26.00){\circle*{0.00}}
\put(47.00,26.00){\circle*{0.00}}
\put(48.00,25.00){\circle*{0.00}}
\put(50.00,25.00){\circle*{0.00}}
\put(49.00,24.00){\circle*{0.00}}
\put(47.00,24.00){\circle*{0.00}}
\put(46.00,23.00){\circle*{0.00}}
\put(48.00,23.00){\circle*{0.00}}
\put(50.00,23.00){\circle*{0.00}}
\put(49.00,22.00){\circle*{0.00}}
\put(47.00,22.00){\circle*{0.00}}
\put(45.00,22.00){\circle*{0.00}}
\put(46.00,21.00){\circle*{0.00}}
\put(48.00,21.00){\circle*{0.00}}
\put(50.00,21.00){\circle*{0.00}}
\put(44.00,21.00){\circle*{0.00}}
\put(47.02,20.18){\circle*{0.00}}
\put(48.95,20.18){\circle*{0.00}}
\put(51.00,30.00){\circle*{0.00}}
\put(52.00,29.00){\circle*{0.00}}
\put(51.00,28.00){\circle*{0.00}}
\put(52.00,27.00){\circle*{0.00}}
\put(51.00,26.00){\circle*{0.00}}
\put(53.00,26.00){\circle*{0.00}}
\put(52.00,25.00){\circle*{0.00}}
\put(54.00,25.00){\circle*{0.00}}
\put(51.00,24.00){\circle*{0.00}}
\put(53.00,24.00){\circle*{0.00}}
\put(54.00,23.00){\circle*{0.00}}
\put(52.00,23.00){\circle*{0.00}}
\put(51.00,22.00){\circle*{0.00}}
\put(53.00,22.00){\circle*{0.00}}
\put(55.00,22.00){\circle*{0.00}}
\put(54.00,21.00){\circle*{0.00}}
\put(52.00,21.00){\circle*{0.00}}
\put(56.00,21.00){\circle*{0.00}}
\put(52.98,20.18){\circle*{0.00}}
\put(51.05,20.18){\circle*{0.00}}
\put(54.00,68.00){\vector(-1,-2){3.50}}
\put(55.00,68.00){\makebox(0,0)[lc]{Center of projection}}
\put(29.00,65.00){\vector(1,-1){4.00}}
\put(28.50,65.00){\makebox(0,0)[rc]{Minkowski ball}}
\bezier{36}(23.86,35.09)(21.75,31.75)(24.74,28.25)
\bezier{92}(38.25,22.63)(49.82,20.88)(61.40,22.63)
\bezier{60}(24.91,28.07)(28.77,24.39)(38.42,22.63)
\bezier{36}(76.14,35.09)(78.25,31.75)(75.26,28.25)
\bezier{60}(75.09,28.07)(71.23,24.39)(61.58,22.63)
\put(50.00,32.63){\line(0,-1){14.56}}
\put(15.96,31.75){\line(5,-2){4.91}}
\put(24.74,28.25){\line(5,-2){13.68}}
\put(44.04,20.53){\line(5,-2){5.96}}
\put(15.96,31.75){\line(5,2){4.56}}
\bezier{30}(25.61,39.65)(33.16,47.54)(39.30,55.80)
\bezier{26}(34.04,60.00)(34.04,50.35)(25.09,39.30)
\put(25.79,39.82){\line(-6,-5){9.82}}
\put(71.93,47.37){\vector(-3,-2){9.65}}
\put(72.28,47.37){\makebox(0,0)[lc]{Initial surface}}
\put(84.04,31.75){\line(-5,-2){4.91}}
\put(75.26,28.25){\line(-5,-2){13.68}}
\put(84.04,31.75){\line(-5,2){4.56}}
\bezier{30}(74.39,39.65)(66.84,47.54)(60.70,55.80)
\bezier{26}(65.96,60.00)(65.96,50.35)(74.91,39.30)
\put(74.21,39.82){\line(6,-5){9.82}}
\put(55.96,20.53){\line(-5,-2){5.96}}
\thicklines
\bezier{204}(24.67,28.17)(44.83,45.83)(38.50,22.67)
\bezier{204}(75.33,28.17)(55.17,45.83)(61.50,22.67)
\put(20.88,29.79){\line(5,-2){3.86}}
\put(38.42,22.77){\line(5,-2){5.61}}
\bezier{16}(20.53,33.68)(19.47,31.75)(20.53,29.82)
\bezier{36}(44.00,20.50)(50.00,19.65)(56.00,20.50)
\put(44.04,20.70){\line(1,2){5.88}}
\put(49.91,32.46){\line(0,1){27.54}}
\bezier{16}(20.53,33.51)(22.11,34.39)(23.68,35.09)
\put(20.53,33.51){\line(5,6){5.12}}
\put(50.00,60.00){\line(-6,-5){24.21}}
\put(20.53,29.87){\line(1,2){4.87}}
\put(79.12,29.79){\line(-5,-2){3.86}}
\put(61.58,22.77){\line(-5,-2){5.61}}
\bezier{16}(79.47,33.68)(80.53,31.75)(79.47,29.82)
\put(55.96,20.70){\line(-1,2){5.88}}
\bezier{16}(79.47,33.51)(77.89,34.39)(76.32,35.09)
\put(79.47,33.51){\line(-5,6){5.12}}
\put(50.00,60.00){\line(6,-5){24.21}}
\put(79.47,29.87){\line(-1,2){4.87}}
\end{picture}

\medskip
\noindent 
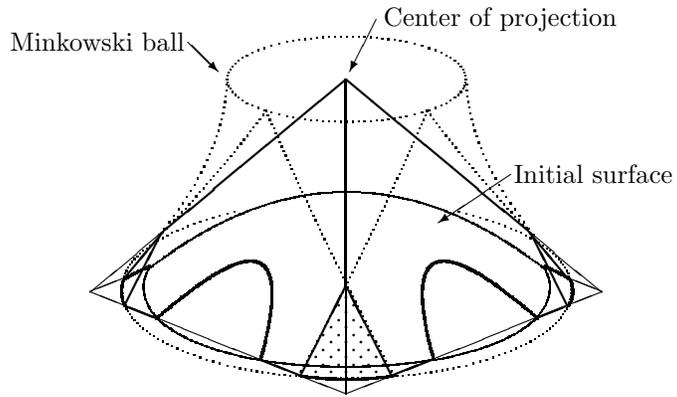
Figure 11: A four-black-hole spacetime in stereoscopic projection centered
on the singularity. As before the physical region should be doubled, with
identifications along the sides of the pyramid. Other identifications
are also possible.\footnote{I am indebted to the group of Bengtsson
at Stockholm, particularly to S\"oren Holst, for clarifying  several features 
of this figure. Also see \cite{I}.} One of the four $\scri$'s 
is shaded by dots. 

\pagebreak

\unitlength 0.90mm
\linethickness{0.4pt}
\begin{picture}(160.00,84.81)
\bezier{34}(14.79,50.00)(14.79,61.27)(24.90,67.68)
\bezier{30}(24.90,67.68)(35.00,72.73)(45.10,67.68)
\bezier{34}(45.10,67.68)(55.21,61.27)(55.21,50.00)
\bezier{34}(14.79,50.00)(14.79,38.73)(24.90,32.32)
\bezier{30}(24.90,32.32)(35.00,27.27)(45.10,32.32)
\bezier{34}(45.10,32.32)(55.21,38.73)(55.21,50.00)
\bezier{52}(15.17,54.03)(19.09,52.97)(22.06,56.65)
\bezier{60}(22.06,56.65)(24.44,60.45)(21.11,64.73)
\bezier{96}(24.17,67.26)(28.80,60.04)(36.13,56.88)
\bezier{116}(36.13,56.88)(45.61,53.16)(54.29,56.09)
\bezier{52}(15.17,45.97)(19.09,47.03)(22.06,43.35)
\bezier{60}(22.06,43.35)(24.44,39.55)(21.11,35.27)
\bezier{96}(24.17,32.74)(28.80,39.96)(36.13,43.12)
\bezier{116}(36.13,43.12)(45.61,46.84)(54.29,43.91)
\put(30.27,50.00){\makebox(0,0)[cc]{S}}
\put(14.69,50.00){\line(1,0){13.54}}
\put(32.30,50.00){\line(1,0){23.01}}
\bezier{52}(41.48,55.18)(40.04,50.00)(41.48,44.67)
\bezier{52}(20.31,54.75)(23.19,50.00)(20.31,45.25)
\put(22.19,51.01){\makebox(0,0)[lb]{$h_1$}}
\put(40.18,51.01){\makebox(0,0)[rb]{$h_3$}}
\put(35.00,10.00){\makebox(0,0)[cc]{(a)}}
\put(120.00,10.00){\makebox(0,0)[cc]{(b)}}
\put(100.00,50.00){\line(1,0){40.00}}
\thicklines
\bezier{36}(139.89,50.00)(140.11,54.44)(142.11,58.89)
\bezier{64}(142.11,58.89)(143.89,64.44)(150.11,72.50)
\bezier{36}(139.89,50.00)(140.11,45.56)(142.11,41.11)
\bezier{64}(142.11,41.11)(143.89,35.56)(150.11,27.40)
\thinlines
\bezier{20}(97.67,58.89)(95.89,64.44)(89.67,72.44)
\bezier{20}(89.67,72.44)(85.22,78.22)(79.89,84.67)
\bezier{20}(97.67,41.11)(95.89,35.56)(89.67,27.56)
\bezier{20}(89.67,27.56)(85.22,21.78)(79.89,15.33)
\thicklines
\bezier{28}(97.89,58.89)(101.22,57.56)(104.56,58.89)
\bezier{76}(104.56,58.89)(111.67,62.67)(119.89,70.00)
\bezier{116}(119.89,70.00)(132.33,62.00)(144.11,70.00)
\bezier{60}(144.11,70.00)(149.00,73.56)(155.89,79.78)
\bezier{28}(97.89,41.11)(101.22,42.44)(104.56,41.11)
\bezier{76}(104.56,41.11)(111.67,37.33)(119.89,30.00)
\bezier{116}(119.89,30.00)(132.33,38.00)(144.11,30.00)
\bezier{60}(144.11,30.00)(149.00,26.44)(155.89,20.22)
\thinlines
\put(97.7,59.00){\line(1,-1){26.44}}
\put(104.78,52.89){\makebox(0,0)[lb]{$h_1$}}
\put(133.67,58.22){\makebox(0,0)[rb]{$h_3$}}
\thicklines
\bezier{80}(97.67,59.11)(101.89,50.00)(97.44,40.89)
\put(150.11,27.56){\line(4,-5){6.04}}
\put(150.11,72.44){\line(4,5){6.04}}
\thinlines
\put(152.78,59.33){\vector(-4,-1){11.11}}
\put(153.00,59.78){\makebox(0,0)[lc]{$\scri_3$}}
\put(90.33,55.56){\vector(4,-1){8.67}}
\put(89.89,56.00){\makebox(0,0)[rc]{$\scri_1$}}
\put(112.5,36.11){\line(1,1){43.70}}
\bezier{6}(156.30,80.19)(158.15,82.59)(160.00,84.81)
\bezier{6}(156.30,19.81)(158.15,17.41)(160.00,15.19)
\end{picture}

\medskip
\noindent 
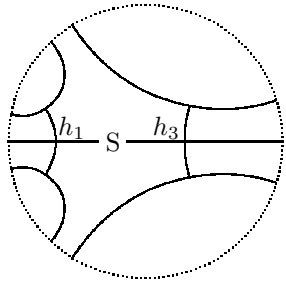
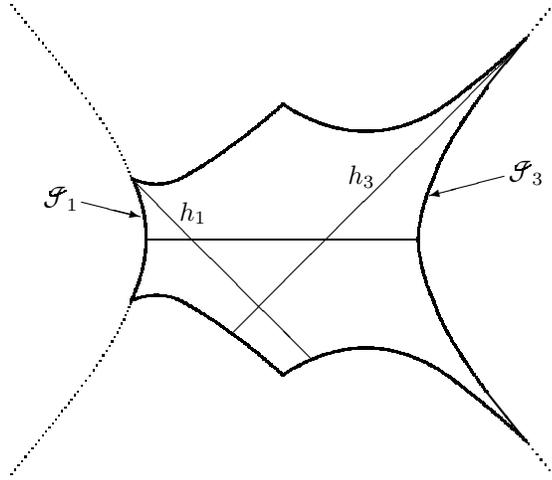
Figure 12: (a) A four-black-hole initial geometry and a line of symmetry, S.

\noindent (b) The spacetime development of the line of symmetry.

\end{document}